\documentclass[11pt]{article}
\usepackage[numbers]{natbib}

\usepackage[title]{appendix}
\usepackage{epstopdf}
\usepackage{amsmath,bm}
\usepackage{empheq}
\usepackage{placeins}
\usepackage{subcaption} 
\usepackage[right]{lineno}
\linenumbers
\usepackage{grffile}

\usepackage{xcolor}

\usepackage{hyperref}

\usepackage[utf8]{inputenc}

\usepackage{nameref,hyperref}
\usepackage{eucal}
\usepackage{amsmath,amssymb}

\usepackage[right]{lineno}

\usepackage{microtype}

\usepackage{changepage}

\usepackage[aboveskip=1pt,labelfont=bf,labelsep=period,singlelinecheck=off]{caption}

\usepackage{lastpage,fancyhdr,graphicx}
\usepackage{epstopdf}

\usepackage{color}

\definecolor{Gray}{gray}{.25}

\usepackage{graphicx}

\usepackage{sidecap}

\usepackage{wrapfig}
\usepackage[fulladjust]{marginnote}

\usepackage{amsfonts,amssymb}

\begin{document}

\author{\noindent
		\begin{tabular}{l}
\textbf{Emanuele De Bono $^1$,  Manuel Collet $^1$} \\
$^1$ Ecole Centrale de Lyon, CNRS, ENTPE, LTDS,\\UMR5513, 69130 Ecully, France\\
\medskip \textbf{emanuele.de-bono@ec-lyon.fr}\\	
\textbf{Morvan Ouisse $^2$} \\
$^2$ SUPMICROTECH, Université de Franche-Comté, CNRS,\\ institut FEMTO-ST, F-25000 Besançon, France.\\
			\end{tabular}
}

\title{The Advection Boundary Law in absence of mean flow: passivity, nonreciprocity and enhanced noise transmission attenuation}

\maketitle

\begin{abstract}\noindent
Sound attenuation along a waveguide is intensively studied for applications ranging from heating and air-conditioning ventilation systems, to aircraft turbofan engines. In particular, the new generation of Ultra-High-By-Pass-Ratio turbofan requires higher attenuation at low frequencies, in less space for liner treatment. This demands to go beyond the classical acoustic liner concepts and overcome their limitations. In this paper, we discuss an unconventional boundary operator, called Advection Boundary Law, which can be artificially synthesized by electroactive means, such as Electroacoustic Resonators. This boundary condition entails nonreciprocal propagation, meanwhile enhancing noise transmission attenuation with respect to purely locally-reacting boundaries, along one sense of propagation. Because of its artificial nature though, its acoustical passivity limits are yet to be defined. In this paper, we provide a thorough numerical study to assess the performances of the Advection Boundary Law, in absence of mean flow. An experimental test-bench validates the numerical outcomes in terms of passivity limits, non-reciprocal propagation and enhanced isolation with respect to local impedance operators. This work provides the guidelines to properly implement the Advection Boundary Law for optimal noise transmission attenuation. Moreover, the tools and criteria provided here can also be employed for the design and characterization of other innovative liners.
\end{abstract}
	

\section{Introduction}\label{sec:intro}

The acoustic problem of interest here, is the noise transmission mitigation in an open duct, by treatment of the parietal walls with a so-called liner. Examples of industrial fields where this problem is particularly felt are the Heating and Ventilation Air-Conditioning Systems (HVAC) and the turbofan aircraft engines. The new generation of Ultra-High-By-Pass-Ratio (UHBR) turbofans, in order to comply with
the significant restrictions on fuel consumptions and pollutant emissions, present larger diameter, lower number of blades and rotational speed and a shorter nacelle. These characteristics conflict with the equally restrictive regulations on noise pollution, as the noise signature is shifted toward lower frequencies, which are much more challenging to be mitigated. The acoustic liner technology applied nowadays for noise transmission attenuation at the inlet and outlet portions of turbofan engines is the so-called Single or Multi-Degree-of-Freedom liner, whose working principle relates to the quarter-wavelength resonance, and demands larger thicknesses to target lower frequencies. They are made of a closed honeycomb structure and a perforated plate which is used to provide the dissipative effect, to add mass in order to decrease the resonance frequency, and also to maintain the aerodynamic flow as smooth as possible on the internal wall of the nacelle. As the honeycomb structure is impervious, propagation is prevented transversely to the wall, therefore it can be considered as \emph{locally reacting} as long as the incident field wavelength is much larger than the size of the honeycomb cells \cite{ma2020development}.\\
A first interest for active control is the possibility to tune the resonators to different frequencies. Many adaptive Helmholtz resonator solutions have been proposed by varying either the acoustic stiffness (i.e. the cavity as in \cite{hermiller2013morphing}), or the acoustic mass (i.e. the orifice area, as in \cite{esteve2004development}), or combining electroactive membranes with Helmholtz resonators \cite{abbad2018adaptive}, but these techniques tended to present complex structure, excessive weight and high energy consumption \cite{ma2020development}.\\
Active Noise Cancellation (ANC) has provided alternative solutions for achieving higher attenuation levels. From the seminal idea of Olson and May \cite{olson1953electronic}, first active \emph{impedance control} strategies \cite{guicking1984active,galland2005hybrid} proposed an ``active equivalent of the quarter wavelength resonance absorber'' for normal and grazing incidence problems, respectively. The same technique was slightly modified in \cite{betgen2011new}, in the attempt to reproduce the Cremer's liner optimal impedance for the first duct modes pair \cite{cremer1953theory,tester1973optimization}. As such impedance could not be achieved in a broadband sense, this approach remained limited to monotonal applications.\\
These are examples of impedance control achieved through secondary source approaches combined with passive liners, but the collocation of sensor and actuator suggested also another avenue: the modification of the actuator (loudspeaker or else) own mechano-acoustical impedance. The objective shifts from creating a ``quiet zone'' at a certain location, to achieving an optimal impedance on the loudspeaker diaphragm.
This is the Electroacoustic Resonator (ER) idea, which have found various declinations, such as electrical-shunting \citep{fleming2007control}, direct-impedance control \citep{furstoss1997surface} and self-sensing \citep{leo2000self}. In order to overcome the low-flexibility drawback of electrical shunting techniques, minimize the number of sensors, meanwhile avoiding to get involved into the electrical-inductance modelling of the loudspeaker, a pressure-based current-driven architecture proved to achieve the best absorption performances in terms of both bandwidth and tunability \citep{rivet2016broadband}. It employs one or more pressure sensors (microphones) nearby the speaker, and a model-inversion digital algorithm to target the desired impedance by controlling the electrical current in the speaker coil. Compared to classical ANC strategies, the impedance control is conceived to assure the acoustical passivity of the treated boundary, and hence the stability of the control system independently of the external acoustic environment \cite{goodwin2001control}. Despite the physiological time delay of the digital control, which can affect the passivity margins at high frequencies \cite{de2022effect}, such ER strategy has demonstrated its efficiency for both room-modal equalization \citep{rivet2016room} and sound transmission mitigation in waveguides \citep{boulandet2018duct,billon2022flow,billon2021experimental,billon20222d,billon2023smart}. The model-inversion algorithm has also been extended to contemplate nonlinear target dynamics at low excitation levels \cite{DeBono2024,da2023experimental,morell2023control,morell2023nonlinear}.\\
All the afore-mentioned boundary treatments for noise mitigation were conceived in terms of target (locally-reacting) behaviors. In \cite{collet2009active}, for the first time, a boundary operator involving the spatial derivative was targeted by distributed electroacoustic devices. It was the first form of the Advection Boundary Law (ABL), then implemented on ER arrays lining an acoustic waveguide in \cite{KarkarDeBono2019,de2023nonlocal,de2023advection}, where it demonstrated non-reciprocal sound propagation. Non-reciprocal propagation is a highly desirable feature for many physical domains and applications \cite{fleury2015nonreciprocal}. In addition, the non-reciprocity allows the ABL to potentially break through typical constraints on the transmission attenuation of reciprocal media \cite{norris2018integral}. Nevertheless, because of its spatial non-locality, the conceptual categories defining the passivity of a surface impedance (see \cite{rienstra2015fundamentals}) do not apply to the ABL. From that, comes the need to reformulate ad-hoc passivity conditions. Moreover, since the ABL lacks any analogue in nature, the physical interpretation of the ABL performances is not immediate. 
In addressing such points, overlooked in the previous references, is the main motivation and contribution of this manuscript.\\
Section \ref{sec:theoretical conception} introduces the ABL from a theoretical point of view and provides a physical parallel which can help in the interpretation of the ABL performances. We arrive to a general definition of the ABL, composed of a convolution impedance operator $\zeta_{Loc}$, and a convection term proportional to the advection speed $U_b$. In Section \ref{sec:passivity discussion semi-inf domain}, the ABL is analysed in open-field to retrieve the oblique incidence absorption coefficient, as function of $U_b$ and of the angle of the incident field.
In Section \ref{sec:duct modes analysis}, the duct-mode eigenproblem is solved by Finite Elements (FEs) for the first modes propagating in a 2D infinite waveguide lined on both sides by the ABL. A \emph{modal} group velocity on the boundary is defined, which allows to introduce the \emph{modal passivity}, as a relaxed version of the \emph{absolute passivity} criteria.
Moreover, the role played by the group velocity angle at the boundary gives the proper understanding of the physical mechanism behind the enhanced attenuation achieved with the ABL. The scattering performances are computed in Setion \ref{sec:scattering 2D} for a 2D duct in the plane wave regime without flow, and a very good correlation is observed with respect to the duct mode analysis. Then, in Section \ref{sec:scatt 3D}, we simulate a 3D waveguide lined by ERs synthesizing the ABL, confirming its enhanced isolation performances, along with its passivity issues. In Section \ref{sec:scatt 3D}, we also present the control law employed to enforce the ABL on the ERs. The effect of discrete pressure estimations by quasi-localized microphones, as well as the impact of time delay in the control algorithm, are briefly discussed. Finally, in Section \ref{sec:experimental grazing incidence}, experimental results are provided to demonstrate the enhanced noise attenuation performances, the broadband nonreciprocal propagation, and the passivity limits of the ABL. The main novelties of this paper stay into the full characterization of the ABL and in the definition of a unique parameter to establish both passivity and noise isolation performances for grazing incidence problems. The nice physical interpretation provided by such parameter, can have an important impact in future designs of acoustic liners.\\
For the reader to easily keep track of the results along the way, at the end of some sections, we list the main outcomes provided there. The overall achievements and next steps are finally discussed in Section \ref{sec:conclusions}.

\FloatBarrier
	
\section{Theoretical conception}\label{sec:theoretical conception}

\begin{figure}
	\centering
	\includegraphics[width=8.6cm]{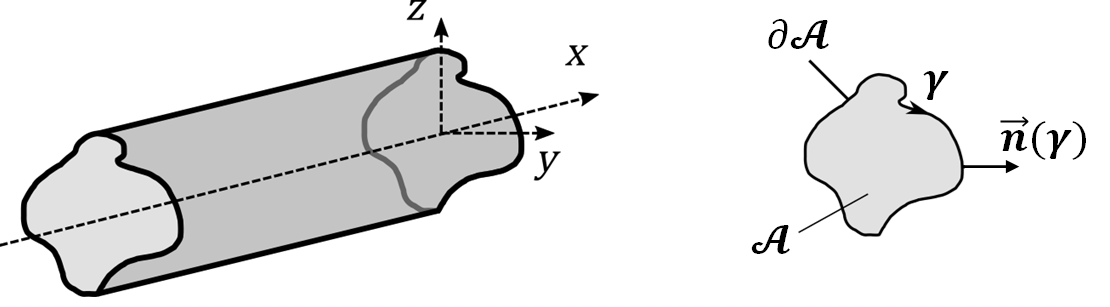}
	\caption{A cylindrical waveguide along coordinate $x$, with cross section $\mathcal{A}$ of arbitrary shape. Left: overview of the waveguide. Right: detail of the cross-section and its contour $\partial\mathcal{A}$. $\vec{n}(\gamma)$ is the local exterior normal at each point of the contour, with tangential coordinate $\gamma$.}
	\label{fig:arbitrary_duct}
\end{figure} 

The greatest difficulty for the parietal treatment of a waveguide is, antonomastically, that it applies on the parietal walls $\partial\mathcal{A}$ (see Fig. \ref{fig:arbitrary_duct}), whereas the noise propagates along the longitudinal axis $x$ which is clearly parallel to $\partial\mathcal{A}$. Such problem is usually referred to as \emph{grazing incidence} problem. C. Bardos, G. Lebeau and J. Rauch \cite{bardos1992sharp} demonstrated that a sufficient condition for the boundary to fully control the wave propagation is that every ray of the acoustic field must interact with the boundary. But in case of the \emph{grazing incidence} problem, there will always be some rays not directly interacting with the boundary, therefore not \emph{controllable}. This is also the reason why the effectiveness of any liner in noise transmission attenuation, degrades if the cross-section area of the waveguide increases, as less number of acoustic rays will directly interact with the boundary.
Nevertheless, even if the grazing incidence problem is not fully controllable, it should still be possible to determine an optimal liner behaviour achieving the maximum attenuation of transmitted noise.\\
Morse \citep{morse1939some}, in 1939, recognized the normal surface impedance as the quantity characterizing the acoustic behaviour of a locally reacting boundary. It is defined as the ratio of Laplace transform of the local sound pressure and the normal velocity: $Z_s(s)=p(s)/v(s)$, where $s$ is the Laplace variable, set to $\mathrm{j}\omega$ (where $\mathrm{j}=\sqrt{-1}$) in the stationary regime. However, a generic boundary might present non-locally reacting, non-linear or even time-variant acoustical response, and in that case the operator describing its acoustical behaviour cannot be reduced to an impedance transfer function.\\
The assumption of locally-reacting behaviour, and its consequent modelling by means of a surface impedance, is common practice in acoustics. Therefore, optimization theories have often considered locally-reacting behaviours of acoustic liners. This is the case for the Cremer's optimal impedance \cite{cremer1953theory}, after retrieved by Tester \cite{tester1973optimization}. Such impedance, formulated in the frequency domain, does not correspond to any real function in time domain (by inverse Fourier transformation), as it does not satisfy the
so-called reality condition \cite{rienstra2015fundamentals}. Attempts to achieve it in a broadband sense \cite{betgen2010comportement} resulted in very large filters, limiting its practical implementation to single tones attenuation \cite{guess1975calculation}. 

\begin{figure}
	\centering
	\includegraphics[width=9.5cm]{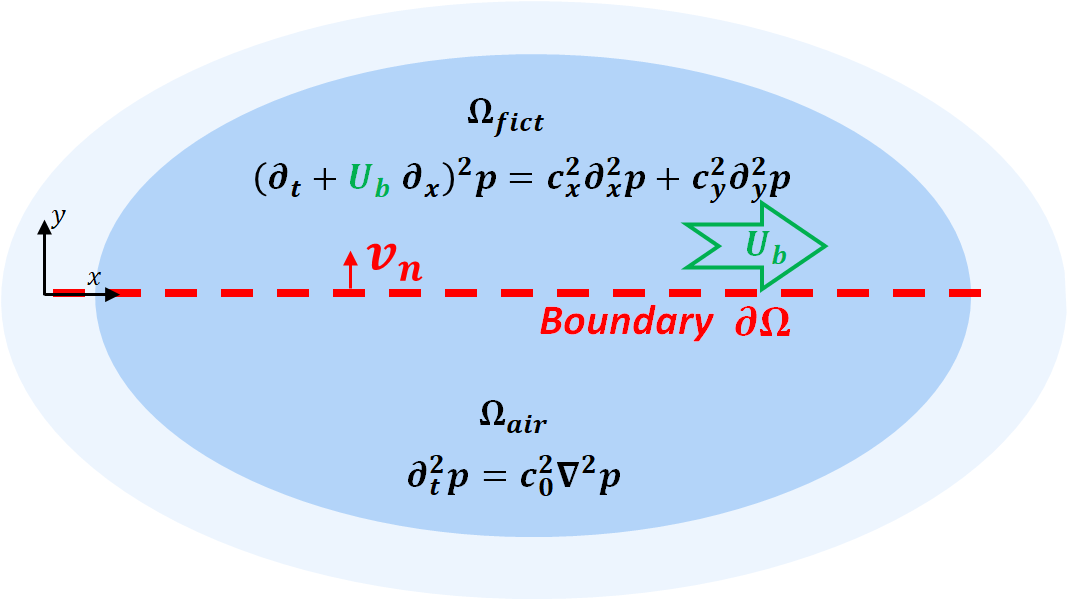}
	\caption{Interface $\partial\Omega$ between two semi-infinite domains: $\Omega_{air}$ and $\Omega_{fict}$. $\Omega_{air}$ is filled with non-convected air, and extends indefinitely toward $\pm x$ and $-y$. $\Omega_{fict}$ extends indefinitely toward $\pm x$ and $+y$, and is an anisotropic acoustic medium characterized by Eq. \eqref{eq:EDB behind wall}.}
	\label{fig:boundary_as_interface_with_FICTICIOUS-DOMAIN}
\end{figure}

To the authors knowledge, general spatial non-local operators have never been targeted for sound transmission attenuation. Nevertheless, Morse himself \cite{morse1939some}, in 1939, introduced the locally reacting surface as a degeneration of a more general interface $\partial\Omega$ (see Fig. \ref{fig:boundary_as_interface_with_FICTICIOUS-DOMAIN}) between two propagative media. The first one $\Omega_{air}$ is characterized by the wave equation in air: 

\begin{equation}\label{eq:wave equation in air}
	c_0^2\nabla^2p = \partial_t^2 p
	\;\;\;\;\textrm{in $\Omega_{air}$},
\end{equation}

with $\nabla$ the Laplacian operator and $c_0$ the sound speed. The other propagative domain, $\Omega_{fict}$ (fictitious) is represented by an anisotropic wave equation, which in 2D reads:

\begin{equation}\label{eq:Morse behind wall}
	c_x^2\partial_x^2 p + c_y^2\partial_y^2p = \partial_t^2 p
	\;\;\;\;\textrm{in $\Omega_{fict}$},
\end{equation}

where $c_x$ and $c_y$ are the phase speeds along the tangential $x$ and normal $y$ coordinates with respect to the boundary $\partial\Omega$.  Observe that in \cite{morse1939some}, Eq. \eqref{eq:Morse behind wall} is written in terms of refractive indices rather than phase speeds. Let us generalize such anisotropic wave equation to take into account a \emph{convection speed} $U_b$ along $x$, in $\Omega_{fict}$:

\begin{equation}\label{eq:EDB behind wall}
	c_x^2\partial_x^2p + c_y^2\partial_y^2 p = (\partial_t+U_b\partial_x)^2 p
	\;\;\;\;\textrm{in $\Omega_{fict}$}.
\end{equation}

Following Morse, a \emph{locally reacting surface} could be interpreted as the interface between air and a domain $\Omega_{fict}$, characterized by Eq. \eqref{eq:EDB behind wall} with $c_x=U_b\ll c_y$, such that both convection and propagation along $x$ can be neglected. This way, Eq. \eqref{eq:EDB behind wall} \emph{degenerates} into a 1D wave equation, where wave propagation in $\Omega_{fict}$ is allowed only along the normal direction $y$ to the surface $\partial\Omega$, with a phase speed equal to $c_y$. The boundary $\partial\Omega$ would then be seen as a locally-reacting surface by $\Omega_{air}$, with characteristic impedance $\rho_{fict}c_y$, with $\rho_{fict}$ the density in $\Omega_{fict}$. For $\Omega_{fict}$ extending to infinity along the $+y$ direction, then the characteristic impedance becomes the surface impedance of the locally-reacting surface $\partial\Omega$. By contemplating complex values of $c_y$ and/or $\rho_{fict}$, complex impedances would be reproduced on the interface $\partial\Omega$.
Usually, non-locally reacting surfaces are attained because $c_x$ is different from zero in Eq. \eqref{eq:EDB behind wall}. It is the case of classical passive non-locally reacting liners (as porous layers), where the $y$-dimension of $\Omega_{fict}$ is bounded by a rigid back wall \cite{ingard2009noise,allard2009propagation}.\\
In the following discussion, $\Omega_{fict}$ will be considered as extending indefinitely from the boundary $\partial\Omega$ toward both coordinate directions ($\pm x$,$+y$), as showed in Fig. \ref{fig:boundary_as_interface_with_FICTICIOUS-DOMAIN} for the 2D case. The definition of a boundary operator corresponding to a rear semi-infinite propagative domain is the so-called Dirichelet-to-Neumann (DtN) mapping \cite{oberai1998implementation}, commonly employed in computational methods for simulating unbounded radiation. The DtN approach is retrieved in \cite{collet2009active} where, by passing through the Fourier space, the pseudo-differential boundary operator (relating sound pressure and its normal derivative), which maps a semi-infinite domain $\Omega_{fict}$ on the interface with $\Omega_{air}$, is computed in case of $\Omega_{fict}$ with same propagation characteristics as $\Omega_{air}$ ($c_y=c_x=c_0$, $U_b=0$).
Following the same steps as \cite{collet2009active}, we can enlarge the pseudo-differential operator presented in \cite{collet2009active} to contemplate an anisotropic and convected propagation in $\Omega_{fict}$ as the one described by Eq. \eqref{eq:EDB behind wall}, and obtain:

\begin{equation}\label{eq:BC with anisotropic}
	\begin{split}
		c_y\partial_y p = -\biggr[\sqrt{(\partial_t+U_b\partial_x)^2 - c_x^2\partial_x^2}\biggr]p \;\;\;\;\textrm{on $\partial\Omega$}.\\
	\end{split}
\end{equation}

In case of $U_b=0$ and $c_x=c_y$, we retrieve the pseudo-differential operator for perfect absorption given in \cite{collet2009active}, while in case of $c_x=U_b=0$, we fall back into the local impedance operator.
Observe that Eq. \eqref{eq:BC with anisotropic} gives the relationship between pressure and its normal derivative, at the interface with a propagative and convected medium. Such relationship is found by imposing the continuity of pressure and normal velocity between the two media \cite{morse1939some,collet2009active}. In real life, the presence of convection and viscosity, would entail a vortex sheet \cite{ingard1959influence,myers1980acoustic,ribner1957reflection} and the continuity of displacement, rather than velocity, at the interface. Nevertheless, as long as we are referring to a fictitious domain $\Omega_{fict}$, this can be assumed inviscid and purely potential, and the continuity of velocity can be maintained as in \cite{keller1955reflection}.\\
Supposing $c_x=0$, Eq. \eqref{eq:BC with anisotropic} degenerates into:

\begin{equation}\label{eq:advection law in dyp}
	\begin{split}
		c_y\partial_y p = -(\partial_t+U_b\partial_x) p \;\;\;\;\textrm{on $\partial\Omega$}.\\
	\end{split}
\end{equation}

Eq. \eqref{eq:advection law in dyp} is the ABL. \textbf{Therefore, we can finally interpret the ABL as the DtN map of a semi-infinite domain $\Omega_{fict}$, characterized by potential wave propagation only along the direction $y$ normal to the boundary (as for locally reacting surfaces), but where such propagation is convected along $x$ with advection speed $U_b$.}
Note that in \cite{collet2009active}, $c_y$ was taken as equal to $c_0$ and Eq. \eqref{eq:advection law in dyp} was not introduced as a degeneration of the general boundary operator (here provided in Eq. \eqref{eq:BC with anisotropic}) mapping a convected anisotropic domain on the boundary. Hence, the introduction of the ABL lacked of a proper physical interpretation.\\
Using the Euler equation of acoustics projected along the $y$-axis (normal to $\partial\Omega$), in absence of mean-flow \cite{filippi1998acoustics}: 

\begin{equation}\label{eq:Euler equation}
		\rho_0\partial_t v_y=-\partial_y p, 
\end{equation}

with $v_y$ the velocity along $y$ (normal to the boundary), Eq. \eqref{eq:advection law in dyp} writes:

\begin{equation}\label{eq:advection law in vy}
	\begin{split}
		\rho_0c_y\partial_t v_y = \partial_t p + U_b\partial_x p \;\;\;\;\textrm{on $\partial\Omega$}.\\
	\end{split}
\end{equation}

Observe that, for $U_b=0$, Eq. \eqref{eq:advection law in vy} retrieves a locally reacting boundary of surface acoustic impedance $Z_{Loc}=\rho_0c_y$. To introduce a general complex local impedance $Z_{Loc}(\mathrm{j}\omega)$, we can define the corresponding differential operator in time domain $Z_{Loc}(\partial_t)$ (same notation as \cite{montseny1998diffusive}), convoluting ($\ast$) the local normal acceleration $\partial_tv_y$.
So, Eq. \eqref{eq:advection law in vy} rewrites:

\begin{equation}\label{eq:advection law in vy with Z}
	\begin{split}
		Z_{Loc}(\partial_t) \ast \partial_tv_y = \partial_t p + U_b\partial_x p \;\;\;\;\textrm{on $\partial\Omega$}.\\
	\end{split}
\end{equation}


In the following, the effects of such BC are investigated first analytically on a semi-infinite domain $\Omega_{air}$, then numerically on a waveguide of infinite and finite lengths, finally experimentally in a duct lined by programmable ERs.

\section{Advection Boundary Law in open field}\label{sec:passivity discussion semi-inf domain}

\begin{figure}
	\centering
	\includegraphics[width=.95\textwidth]{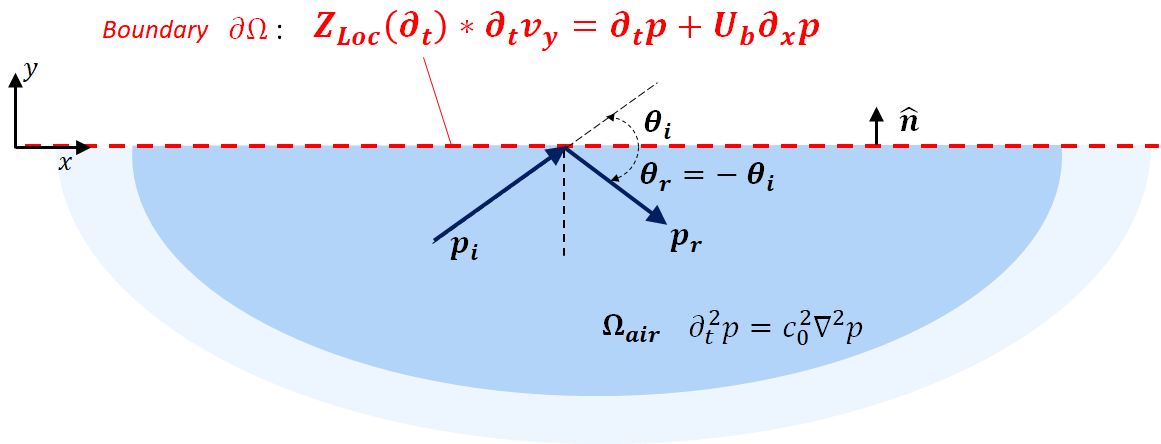}
	\caption{ABL interfacing a semi-infinite domain.}
	\label{fig:boundary_advectionwith_SemiInfDomain}
\end{figure}

As a first case study, we compute the absorption coefficient of the ABL interfacing a semi-infinite air domain (an open field), $\Omega_{air}=[-\infty,\infty]\times[-\infty,0]$, as in Fig. \ref{fig:boundary_advectionwith_SemiInfDomain}. The treated boundary extends on all the $x$ axis.

Assuming a time-harmonic sound field in the usual complex notation ($\mathrm{j}\omega t$), the incident wave can be expressed as:

\begin{equation}
	\label{eq:incident free plane wave}
	\bar{p}_i(t,\omega,x,y)=p_{i0}(\omega)e^{\mathrm{j}\omega t - \mathrm{j}k_0\cos\theta_i x - \mathrm{j}k_0\sin\theta_i y},
\end{equation}

where $\bar{p}_i$ is the complex representation of $p_i=\mathrm{Re}\{\bar{p}_i\}$, $k_0=\omega/c_0$ is the wavenumber of a plane wave, $c_0$ is the speed of sound, and $\theta_i$ is the incident angle of the plane wave on the treated boundary. The reflected wave field is supposed to respect the classical Snell-Descartes law of refraction, according to which the reflected plane wave propagates with a specular angle with respect to the incident one, i.e. $\theta_r=-\theta_i$. The presence of a transport at the boundary gives no reason to modify this assumption, in an analogous way to the case of air-flow in the acoustic domain \cite{ingard1959influence}, or to the case of interface with a convected propagative medium \cite{ribner1957reflection}.\\
Hence, the complex reflected wave from an ABL can be written as:

\begin{equation}
	\label{eq:reflected free plane wave}
	\bar{p}_r(t,\omega,x,y)=R(\mathrm{j}\omega)p_{i0}e^{\mathrm{j}\omega t - \mathrm{j}k_0\cos\theta_i x + \mathrm{j}k_0\sin\theta_i y},
\end{equation}

with $R$ the reflection coefficient at the oblique incidence $\theta_i$.
The acoustic velocity $v_y$ normal to the boundary is obtained by the Euler equation of acoustics projected along $y$ (Eq. \eqref{eq:Euler equation}), with $p=p_i+p_r$. Replacing Eq.s \eqref{eq:incident free plane wave} and \eqref{eq:reflected free plane wave} in $p_i$ and $p_r$ respectively, we find the normal complex velocity on the boundary $y=0$:

\begin{equation}
	\label{eq:velocity from Euler from incident and reflected free plane waves}
	\bar{v}_y(t,x)=\frac{\sin\theta_i}{\rho_0c_0} p_{i0} \biggr(1 - R(\mathrm{j}\omega)\biggr) e^{\mathrm{j}\omega t - \mathrm{j}k_0\cos\theta_i x}.
\end{equation}

Also, the ABL of Eq. \eqref{eq:advection law in vy with Z} can be applied to the total pressure $p=p_i + p_r$ to give:

\begin{equation}
	\label{eq:velocity at advective BC from incident and reflected free plane waves}
	\bar{v}_y(t,x)=\frac{p_{i0}}{Z_{Loc}(\mathrm{j}\omega)}\biggr(1 + M_b\cos\theta_i \biggr) \biggr(1 + R(\mathrm{j}\omega)\biggr)e^{\mathrm{j}\omega t - \mathrm{j}k_0\cos\theta_i x},
\end{equation}

where $M_b=U_b/c_0$. Equating Eq. \eqref{eq:velocity from Euler from incident and reflected free plane waves} and \eqref{eq:velocity at advective BC from incident and reflected free plane waves}, we find the reflection coefficient:

\begin{equation}\label{eq:refl coeff advection law}
	R(\omega) =
	\frac{ 1 - \biggr(1-M_b\cos\theta_i\biggr)\eta_{Loc}(\mathrm{j}\omega)/\sin\theta_i  }
    	 { 1 + \biggr(1-M_b\cos\theta_i\biggr)\eta_{Loc}(\mathrm{j}\omega)/\sin\theta_i},
\end{equation}

where $\eta_{Loc}(\mathrm{j}\omega)=\rho_0c_0/Z_{Loc}(\mathrm{j}\omega)$ is the normalized local mobility. Observe that for $M_b=0$, the reflection coefficient of classical locally-reacting surfaces is retrieved. Eq. \eqref{eq:refl coeff advection law} suggests the possibility to define an \emph{effective} normalized mobility $\eta_{eff}(\mathrm{j}\omega,M_b,\theta_i)=\biggr(1 - M_b\cos\theta_i\biggr)\eta_{Loc}(\mathrm{j}\omega)$, which is equivalent to the ABL operator for the far-field reflection from an infinite boundary $\partial\Omega$. Observe that $\eta_{eff}$ depends also on $M_b$ and $\theta_i$. In particular, it is interesting to notice that for $M_b=-1$, if $\theta_i\rightarrow0$ then $\eta_{eff}\rightarrow 2 \eta_{Loc}$, whereas if $\theta_i\rightarrow\pi$ then $\eta_{eff}\rightarrow0$. This result preliminarily demonstrates the non-reciprocal propagation achieved by the ABL in grazing incidence, which is treated in the next sections.\\
Based on $\eta_{eff}$, we can write the absorption coefficient:

\begin{equation}\label{eq:absorption coeff advection law}
	\alpha(\omega) =
	\frac{ 4\; \mathrm{Re}\{ \eta_{eff}(\mathrm{j}\omega,\theta_i,M_b)/\sin\theta_i \} }
	{\biggr|1 + \eta_{eff}(\mathrm{j}\omega,\theta_i,M_b)/\sin\theta_i\biggr|^2}.
\end{equation}
	
From Eq. \eqref{eq:absorption coeff advection law}, we can apply the classical passivity condition for locally-reacting boundaries \cite{rienstra2015fundamentals} to $\eta_{eff}(\mathrm{j}\omega,\theta_i,M_b)$:

\begin{equation}\label{eq:passivity condition semi-inf domain}
\mathrm{Re}\biggr\{\eta_{eff}(\mathrm{j}\omega,\theta_i,M_b)\biggr\}\geq 0 \quad \mathrm{i.e.} \quad
\mathrm{Re}\biggr\{\eta_{Loc}(\mathrm{j}\omega)\biggr\}\biggr(1-M_b\cos\theta_i\biggr) \geq 0.
\end{equation}

Eq. \eqref{eq:passivity condition semi-inf domain} is valid as long as $\mathrm{Re}\{\eta_{Loc}(\mathrm{j}\omega)\}\geq 0$ (the local impedance operator should be passive) and $M_b\leq1/\cos\theta_i$. For the passivity to hold independently of the angle of incidence, it must be $M_b\leq1$. Such acoustical passivity condition signifies that acoustic energy \emph{enters} the boundary, rather than being radiated from it. Let us write the acoustic intensity \cite{filippi1998acoustics} normal to the boundary, $I_y$:

	\begin{equation}\label{eq:normal acoustic intensity open field}
		I_y(\omega)=
		\frac{1}{2}\mathrm{Re}\{\bar{p}^*(\omega,x)\bar{v}_y(\omega,x)\}=
		\biggr(1 - |R(\omega)|^2 \biggr)\frac{|p_{i0}(\omega)|^2}{2\rho_0c_0}\sin\theta_i =
		\alpha(\omega)\frac{|p_{i0}(\omega)|^2}{2\rho_0c_0}\sin\theta_i=\alpha(\omega)I_{i,y}(\omega),
	\end{equation}

where $I_{i,y}(\omega)=\frac{|p_{i0}(\omega)|^2}{2\rho_0c_0}\sin\theta_i$ is the component along $y$ of the incident acoustic intensity, and the superscript $^*$ indicates the complex conjugate. Therefore, for a given incident field, the normal component of the acoustic intensity gives the absorption at the boundary. 

\begin{figure}
	\centering
	\includegraphics[width=.65\textwidth]{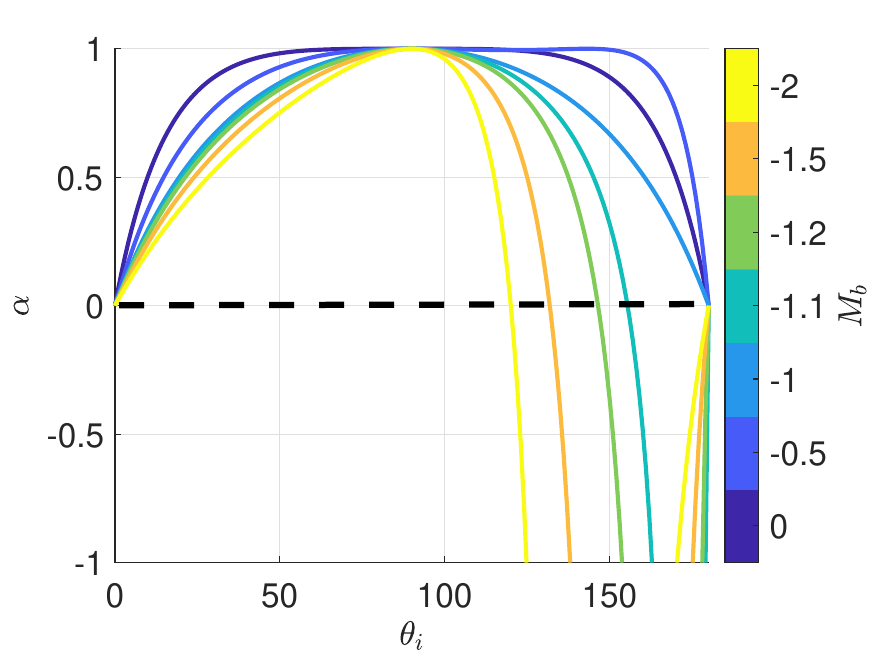}
	\caption{ABL absorption coefficient versus $\theta_i$, in open field, for $\eta_{Loc}=1$ and various $M_b\leq0$.}
	\label{fig:alphaVSthetai_varyingMb}
\end{figure}

The ABL absorption coefficient versus the angle of incidence is plotted in Fig. \ref{fig:alphaVSthetai_varyingMb} for $\eta_{Loc}=1$ and different values of $M_b<0$. Notice that the range of angles of incidence $\theta_i$ for which $\alpha<0$, enlarges as $|M_b|$ is increased above 1. Moreover, such loss of acoustical passivity for $M_b<-1$, happens only for $\pi/2<\theta_i<\pi$, meaning that the ABL is non-passive only for incident sound fields coming from the right side of Fig. \ref{fig:boundary_advectionwith_SemiInfDomain}, that is, for incident waves with $\mathrm{sgn}(k_x)=\mathrm{sgn}(M_b)$. The dependence upon the angle of incidence of ABL acoustical passivity is another unique feature of the ABL with respect to classical liners. This angle-of-incidence dependency of ABL acoustical passivity manifests in a duct-mode dependent stability, which is the subject of the next section.\\ 

Below, the main outcomes of Section \ref{sec:passivity discussion semi-inf domain}:

\begin{itemize}
\item \textbf{In open field, the ABL passivity conditions are $\eta_{Loc}\geq0$ and $|M_b|\leq1$.}
\item \textbf{In open field, the normal acoustic intensity at the boundary is related to the absorption coefficient of the boundary.}
\item \textbf{In open field, ABL acoustical passivity depends upon the angle of the incident field.}
\item \textbf{In open field, the passivity loss is larger for higher absolute values of $M_b<-1$.}
\end{itemize}


\FloatBarrier

\section{Duct modes analysis in 2D waveguide}\label{sec:duct modes analysis}

\begin{figure}
	\centering
	\includegraphics[width=8.6cm]{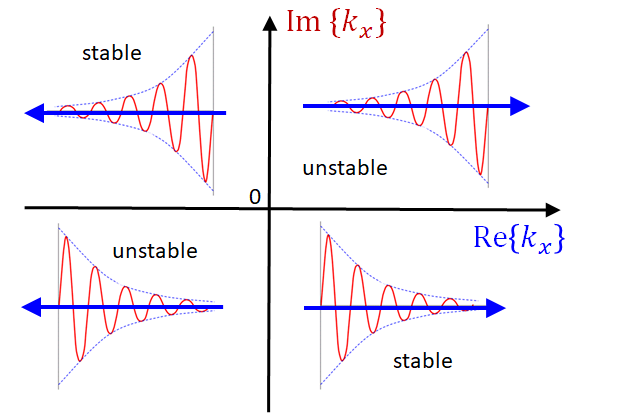}
	\caption{Stability regions of duct-modes in the ($\mathrm{Re}\{k_x\}$,$\mathrm{Im}\{k_x\}$)-plane.}
	\label{fig:duct_modes_stability_regions}
\end{figure}


After having defined the passivity condition of the ABL on a semi-infinite domain, let us investigate the passivity and attenuation performances into an acoustic waveguide starting from the duct mode analysis. Duct modes are fundamental to understand the propagation characteristics in a waveguide. The general formulation of the duct-mode eigen-problem is provided in Appendix \ref{app:Duct modes problem formulation}, along with the special treatment reserved to the ABL for the FE numerical resolution of our eigen-problem. The FE mesh has been built sufficiently fine to have large number of elements in the cross section and accurately resolve for each duct-mode shape of interest.
We consider a 2D duct of section width $h=0.05$ m, with both upper and lower walls lined by the ABL. According to the assumption of duct mode eigen-solution $\bar{p}_m(t,\omega,x,y) = A_m \psi_{m}(y,\omega) e^{\mathrm{j}\omega t - \mathrm{j}k_{x,m}(\omega)x}$, the duct mode analysis consists in computing the duct-mode eigenvalues ($k_{x,m}$) and eigenvectors ($\psi_{m}$), while $A_m$ can be normalized at will. The duct-mode representation of the acoustic field, gives the occasion to define \emph{modal acoustic intensities} and \emph{modal group velocities}. In particular, the \emph{local} modal acoustic intensity vector is given by: 

\begin{equation}\label{eq:local acoustic intensity vector}
 \vec{I}_m(x,y,\omega) = \frac{1}{2} \mathrm{Re}\{\bar{p}^*_m(t,\omega,x,y)\vec{v}_m(t,\omega,x,y)\},
\end{equation}

where the superscript $*$ indicates the complex conjugate, and $\vec{v}_m$ is the modal acoustic velocity, related to the modal acoustic pressure $\bar{p}_m$ by the Euler equation of acoustics  $-\rho_0\mathrm{j}\omega \vec{v}_m = \vec{\nabla} \bar{p}_m$, where $\vec{\nabla}$ is the gradient operator. We can then compute the $x$ and $y$ components of $\vec{I}_m$:

\begin{subequations}\label{local acoustic intensity components}
\begin{equation}\label{Ixm}
I_{x,m}(x,y,\omega) = e^{2\mathrm{Im}\{k_{x,m}x\}}\frac{1}{2\rho_0c_0} \frac{\mathrm{Re}\{k_{x,m}\}}{k_0}|\psi_m(y,\omega)|^2
\end{equation}
\begin{equation}\label{Iym}
	I_{y,m}(x,y,\omega) = e^{2\mathrm{Im}\{k_{x,m}x\}}
	\frac{1}{2\rho_0c_0}
	\mathrm{Re}\biggr\{\psi_m^*(y,\omega)\partial_y\psi_m(y,\omega)\biggr\}
\end{equation}
\end{subequations}

We can now define the \emph{average} acoustic intensity vector on the duct cross section:

\begin{equation}\label{eq:average acoustic intensity vector}
	\vec{I}_{m,Ave}(x,\omega)=\frac{1}{h}\int_{0}^{h} \vec{I}_m(x,y,\omega)dy.
\end{equation}

It is easy to verify that, for symmetric duct modes (for which $\psi_m(h,\omega)=\psi_m(0,\omega)$), we get $\int_{0}^{h}I_{y,m}(x,y,\omega)dy=0$. So:

\begin{equation}\label{eq:average acoustic intensity vector expression}
\vec{I}_{m,Ave}(\omega)=\frac{1}{h}\biggr(\int_{0}^{h} I_{x,m}(y,\omega)dy\biggr)\vec{x}
=e^{2\mathrm{Im}\{k_{x,m}\}x}\frac{1}{2h\rho_0c_0}
\frac{\mathrm{Re}\{k_{x,m}\}}{k_0}\biggr(\int_{0}^{h} |\psi_m(y,\omega)|^2 dy \biggr) \vec{x},
\end{equation}

where $\vec{x}$ is the unit vector along $x$.\\
We can now define the local modal group velocity as:

\begin{equation}\label{eq:local group velocity}
\vec{c}_m(y,\omega) = \frac{\vec{I}_m(x,y,\omega)}{E_{m,Ave}(x,\omega)},
\end{equation}

where $E_{m,Ave}(x,\omega)$ is the average modal kinetic energy, defined as:

\begin{equation}
\begin{split}
E_{m,Ave}(x,\omega)&=\frac{\rho_0}{2h}\int_{0}^{h}\vec{v}^*_m\cdot\vec{v}_mdy\\
&=\frac{e^{2\mathrm{Im}\{k_{x,m}x\}}}{2h\rho_0c_0^2} \biggr(\frac{|k_{x,m}|^2}{k_0^2} \int_{0}^{h} |\psi_m(y,\omega)|^2 dy + \frac{1}{k_0^2}\int_{0}^{h}|\partial_y\psi_m(y,\omega)|^2 dy \biggr).
\end{split}
\end{equation}

We can then compute the average modal group velocity:

\begin{equation}\label{eq:qverage group velocity}
\begin{split}
\vec{c}_{m,Ave}(\omega)
&= \frac{1}{h}\int_{0}^{h}\vec{c}_m(x,y,\omega)dy= \frac{1}{h}\frac{\int_{0}^{h}\vec{I}_m(x,y,\omega)dy}{E_{m,Ave}(x,\omega)}\\
&= c_0\frac{\mathrm{Re}\{k_{x,m}(\omega)\}}{k_0}\Biggr(\frac{\int_{0}^{h}|\psi_m(y,\omega)|^2dy}{\frac{|k_{x,m}|^2}{k_0^2}\int_{0}^{h}|\psi_m(y,\omega)|^2dy + \frac{1}{k_0^2}\int_{0}^{h}\partial_y|\psi(y,\omega)|^2}\Biggr)\vec{x},
\end{split}
\end{equation}

Observe that neither the local or the average modal group velocities depend upon $x$.
From the average modal group velocity expression, we can deduce that each duct mode propagates along $x$ with a sign given by $\mathrm{Re}\{k_{x,m}\}$, i.e. Re$\{k_{x,m}\}$$>0$ means a $+x$ direction of propagation, and vice-versa. The $\mathrm{Im}\{k_{x,m}\}$ instead, gives the attenuation (or amplification) rate of the modal acoustic intensity along the duct mode $x$-propagation, as it can be seen from the Eq. \eqref{eq:average acoustic intensity vector expression}. The regions of duct-mode stability are illustrated in Fig. \ref{fig:duct_modes_stability_regions}, for clarity. However, we are interested in defining a unique dimensionless quantity able to characterize both the attenuation and stability of a duct mode. Inspired by the work of Rice \cite{rice1979modal}, we propose to consider the propagation angle of the local modal group velocity at the boundary, given by:

\begin{equation}\label{eq:modal group velocity propagation angle at the boundary}
\theta_{b,m}(\omega) = \mathrm{atan}\biggr(\frac{c_{n,m}(\omega,y_b)}{c_{x,m}(\omega,y_b)}\biggr),
\end{equation}

where $c_{n,m}$ is the local modal group velocity component along the normal $\vec{n}$ to the boundary, and $y_b$ is the value of the $y$ coordinate at the boundary. Clearly, a dissipative liner entails acoustic intensity that enters the boundary, i.e. a positive $c_{n,m}(\omega,y_b)$ and a $0<\theta_{b,m}<\pi$. 
Therefore, we propose to define the following \emph{absolute} acoustical passivity criteria of a generic BC for in-duct grazing-incidence problems:

\begin{equation}\label{eq:absolute acoustical passivity grazing incidence}
\sin\theta_{b,m}(\omega) \geq 0, \quad \forall \; \omega>0\ \; \mathrm{and} \;\;  \forall m\in\mathbb{Z}^+.
\end{equation}

Such \emph{absolute} passivity criteria could be relaxed to introduce a more general \emph{modal} acoustic passivity criteria:

\begin{equation}\label{eq:modal acoustical passivity grazing incidence}
\sin\theta_{b,m}(\omega) \geq 0 \quad \forall \; \omega>0\ \; \text{and certain} \;\; m\in\mathbb{Z}^+,
\end{equation}

Such quantity $\sin\theta_{b,m}$ very well correlates also with the attenuation levels achieved by the ABL, as it will be showed in the following. Observe that this quantity differs from the modal propagation angle considered by Rice \cite{rice1979modal} to correlate with the acoustic liner performances. In \cite{rice1979modal}, the modal propagation angle at the boundary was computed from the wavenumber, and not from $\vec{c}_{m}$. Indeed, the group velocity was considered only in case of air-flow in the duct. Moreover, he proposed a geometric approach employing the open-field reflection coefficient computed for an incident angle equal to such modal propagation angle, to estimate the attenuation rate along the duct, achieving good approximation only for nearly hard walls (locally-reacting liners with $\eta_{Loc}\approx0$). Moreover, the separation between incident and reflected fields cannot be operated in a duct-mode analysis, therefore such open-field reflection coefficient actually provides very poor estimations of the attenuation rates for general BCs of interest.\\

The solutions, both in terms of wavenumbers $k_{x,m}$ and mode-shapes $\psi_{m}(y)$ reported here, are computed for a 2D waveguide with cross section width $h=0.05$ m (to conform with the experimental test-rig of Section \ref{sec:experimental grazing incidence}), lined on both sides by our ABL. The results will be accompanied by the plots of $\sin\theta_{b,m}$ to demonstrate the perfect correlation of duct-mode stability with the modal passivity criteria of Eq. \eqref{eq:modal acoustical passivity grazing incidence}, and the good correlation with the attenuation rate given by Im$\{k_{x,m}\}$.

\subsection{Real local impedance $\zeta_{Loc}$}\label{sec:duct modes real impedance}
\begin{figure}[ht!]
		\centering
		\includegraphics[width=0.7\textwidth]{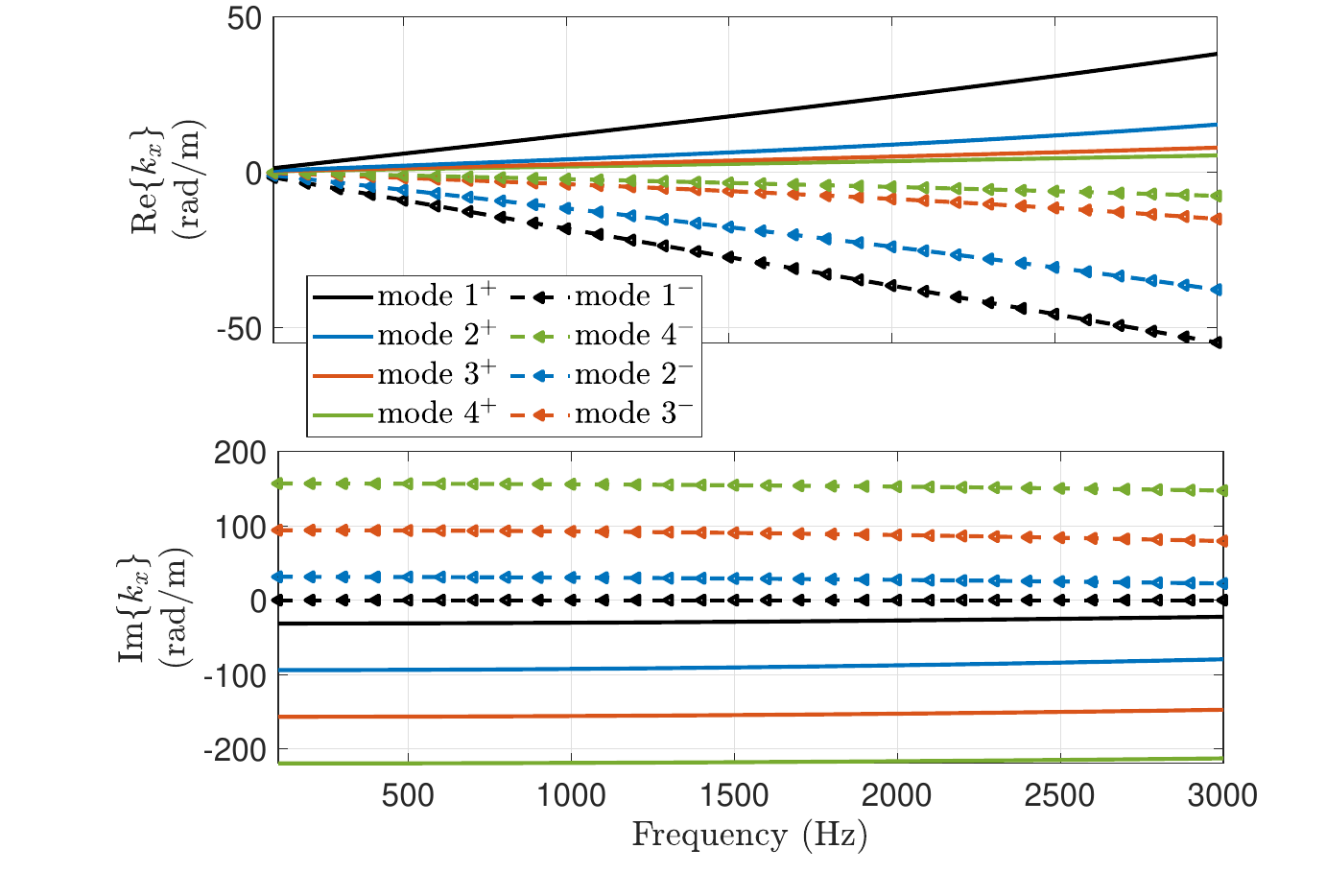}
		\centering
		\label{fig:K_Omega_plane_NonLocalBC_RatRho0c0_Kat0_Mat0_-1cac0}
	\caption{Dispersion plots for the wavenumbers relative to the first four duct modes propagating in both senses, in case of boundary advection law with $\eta_{Loc}=1$ and $M_b=-1$.}
	\label{fig:NonLocalBCresistive_ca-1_KvsOmega}
\end{figure}

\begin{figure}[ht!]
	\centering
	\begin{subfigure}[ht!]{0.45\textwidth}
		\centering
	\includegraphics[width=\textwidth]{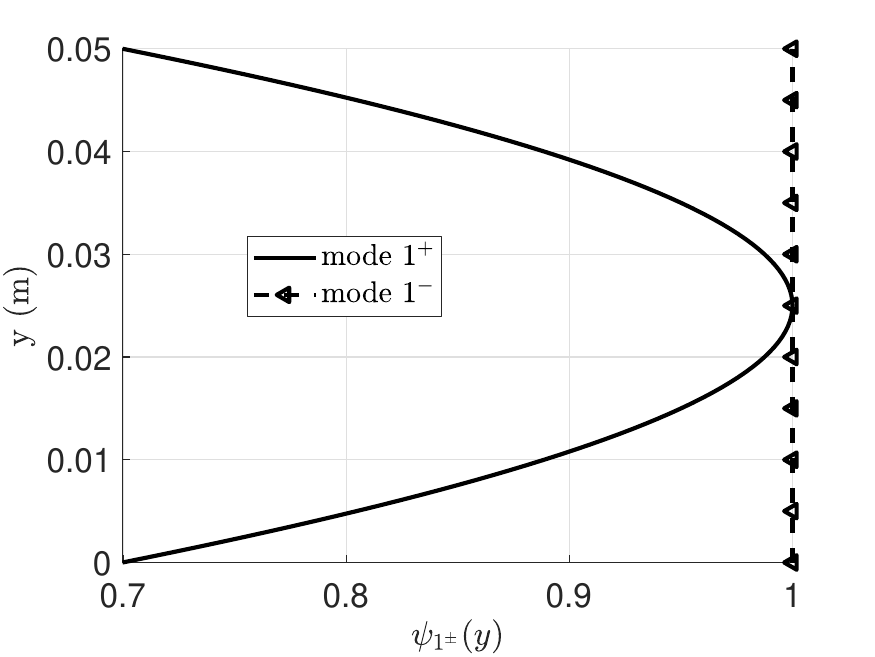}\label{fig:mode1pm shapes}
\caption{}
	\end{subfigure}
	\hspace{0.5 cm}
	\begin{subfigure}[ht!]{0.45\textwidth}
		\centering
	\includegraphics[width=\textwidth]{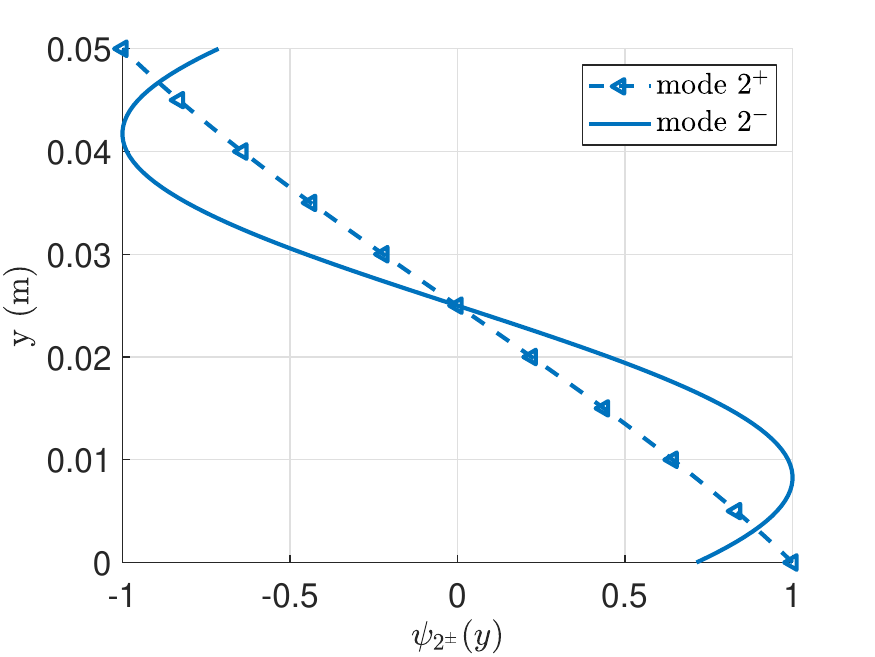}\label{fig:mode2pm shapes}
\caption{}
	\end{subfigure}
	\begin{subfigure}[ht!]{0.45\textwidth}
	\centering
	\includegraphics[width=\textwidth]{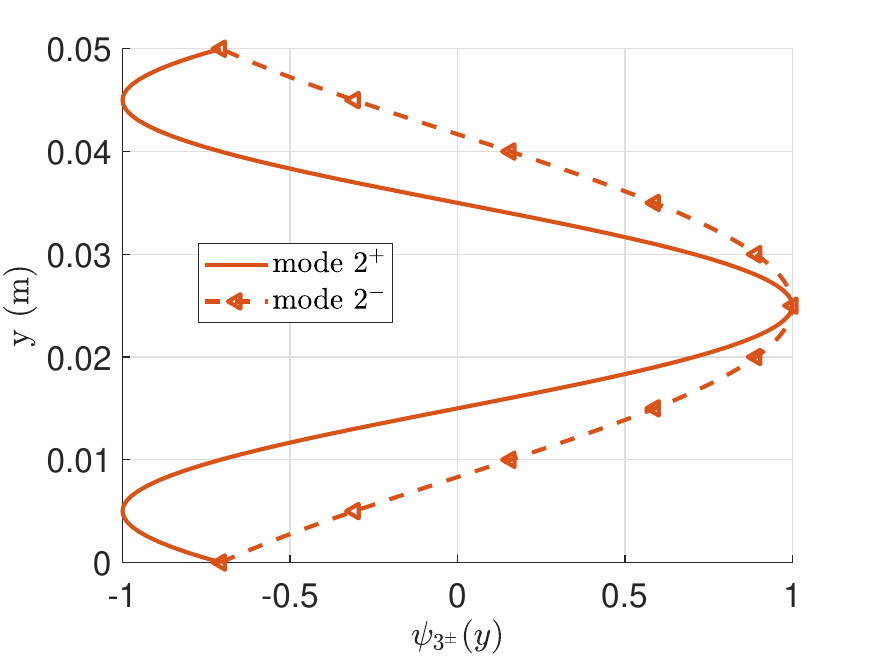}\label{fig:mode3pm shapes}
\caption{}
\end{subfigure}
\hspace{0.5 cm}
\begin{subfigure}[ht!]{0.45\textwidth}
\centering
\includegraphics[width=\textwidth]{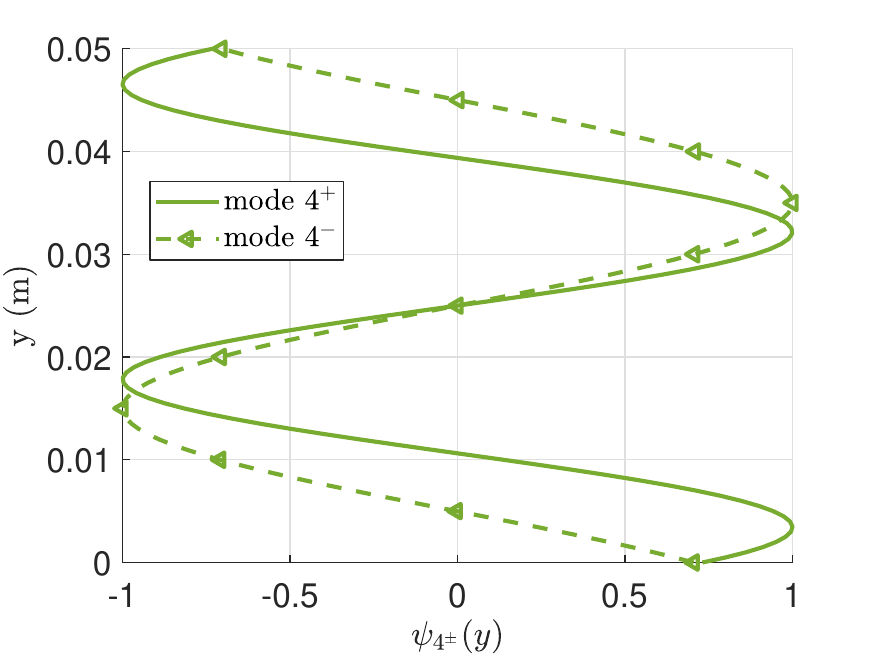}\label{fig:mode4pm shapes}
\caption{}
\end{subfigure}
	\caption{First four duct mode-shapes $\psi_{m}(y)$, propagating toward positive and negative $x$ direction, normalized with respect to the maximum value, for ABL treated boundaries with $\eta_{Loc}=1$ and $M_b=-1$, at 500 Hz.}
\label{fig:NonLocalBCresistive_ca-1_Modes}
\end{figure}


In this Section, the local impedance, and hence the local normalized mobility $\eta_{Loc}$, is considered as purely real.\\
In Fig.s \ref{fig:NonLocalBCresistive_ca-1_KvsOmega} and \ref{fig:NonLocalBCresistive_ca-1_Modes} the first eight solutions in terms of wavenumbers and corresponding duct modes respectively, are plotted. The frequency span is limited between $150$ and $3000$ Hz to focus on the same frequency range as the experimental results.
It is evident that the mode-shapes propagating towards $+x$ present a shorter wavelength along $y$ with respect to those propagating toward $-x$. Moreover, one can notice that mode $1^+$ is attenuated (Im$\{k_{x,1^+}\}<0$), while mode $1^-$ is a plane wave ($\psi_1^-=1$, $k_{x,1^-}=-k_0$). This demonstrates \emph{the breaking of the reciprocity principle} \cite{fleury:2015} in the plane wave regime, as it will be clearer in the following.\\

\begin{figure}[ht!]
	\centering
	\begin{subfigure}[ht!]{0.45\textwidth}
		\centering
		\includegraphics[width=\textwidth]{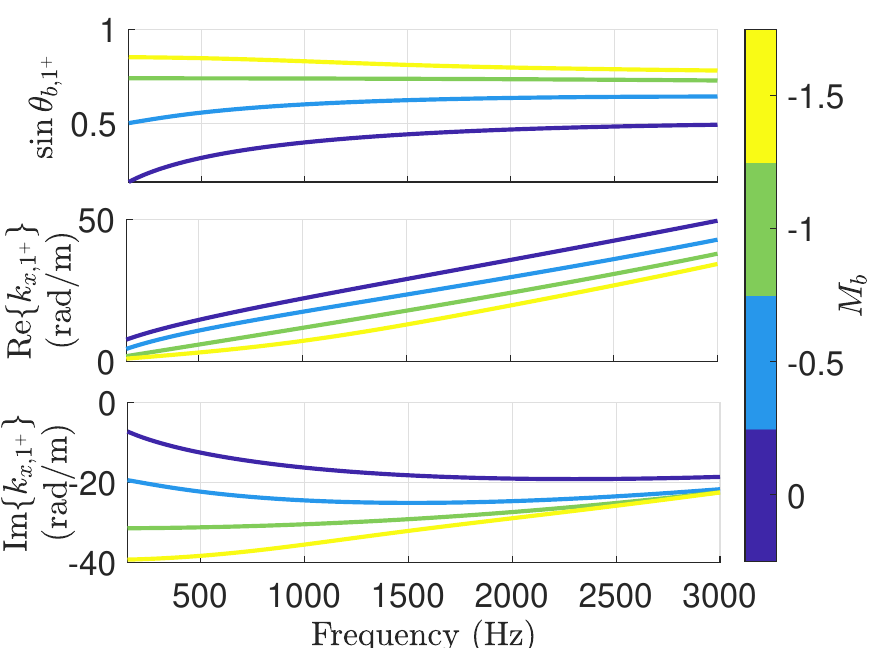}
		\caption{}
		\label{fig:cgnANDReImKxVSfreq_mode9_varyingMb_PurelyRealEtaLoc}
	\end{subfigure}
	\hspace{0.5 cm}
	\begin{subfigure}[ht!]{0.45\textwidth}
		\centering
		\includegraphics[width=\textwidth]{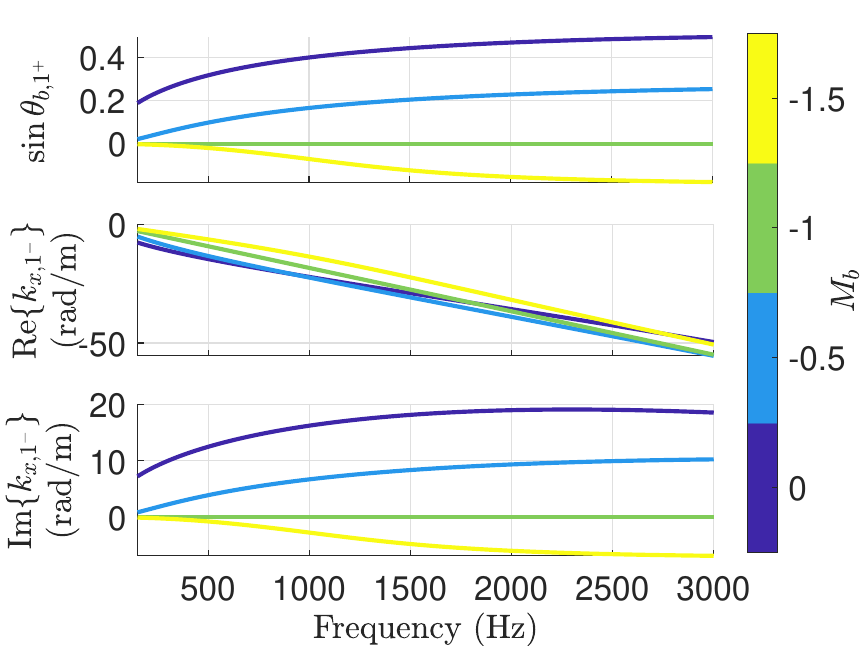}
		\caption{}
		\label{fig:cgnANDReImKxVSfreq_mode1_varyingMb_PurelyRealEtaLoc}
	\end{subfigure}
	\caption{Spectra of $\sin\theta_{b,m}$, $\mathrm{Re}\{k_{x,m}\}$ and $\mathrm{Im}\{k_{x,m}\}$, with $\eta_{Loc}=1$ and varying $M_b<0$, for mode $m=1^+$ \textbf{(a)} and $m=1^-$ \textbf{(b)}.}
	\label{fig:cgnANDReImKxVSfreq_mode9and1}
\end{figure}


In this paper we study just the first forward and backward propagating mode ($1^+$ and $1^-$), as we are interested in the isolation performances in the plane wave regime of a rigid duct. Indeed, the first modes are also the least attenuated ones, therefore mostly ruling the noise transmission when the liner is applied in a segment of a rigid duct \cite{munjal1990theory,cremer1953theory}. Fig. \ref{fig:cgnANDReImKxVSfreq_mode9and1} shows the frequency plots of $\sin\theta_{b,m}$, $\mathrm{Re}\{k_{x,m}\}$ and $\mathrm{Im}\{k_{x,m}\}$, for modes $m=1^+$ and $m=1^-$. Looking at Fig. \ref{fig:cgnANDReImKxVSfreq_mode1_varyingMb_PurelyRealEtaLoc}, we observe that for $M_b=-1$, mode $1^-$ becomes a plane wave, while for $M_b<-1$ we have non-stable duct mode propagation, confirmed by a $\sin\theta_{b,1^-}<0$. Notice that the attenuation rate ($\mathrm{Im}\{k_{x,1^-}\}$) follows the same trend as $\sin\theta_{b,1^-}<0$ with $M_b$, and also with frequency. Looking at Fig. \ref{fig:cgnANDReImKxVSfreq_mode9_varyingMb_PurelyRealEtaLoc}, notice the monotonic increase of both $\sin\theta_{b,1^+}$ and $\mathrm{Im}\{k_{x,1^+}\}$ with $|M_b|$, confirming the good correlation between these two quantities, and the higher attenuation performances achievable thanks to the ABL with $M_b<0$ with respect to local impedance operators ($M_b=0$). Nevertheless, at high frequencies, $\mathrm{Im}\{k_{x,1^+}\}$ for $M_b=-1.5$ seems to almost coalesce with $M_b=-1$ and $M_b=-0.5$, which is not the same for $\sin\theta_{b,1^+}$. We can then state that the correlation between $\sin\theta_{b,1^+}$ and the attenuation rate is very high at lower frequencies. 

\begin{figure}[ht!]
	\centering
	\begin{subfigure}[ht!]{0.45\textwidth}
		\centering
		\includegraphics[width=\textwidth]{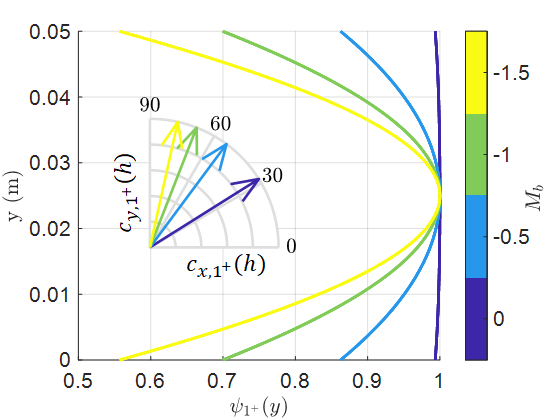}
		\caption{}
		\label{fig:mode1plus_500Hz_varyingNegMb_withCompass}
	\end{subfigure}
	\hspace{0.5 cm}
	\begin{subfigure}[ht!]{0.45\textwidth}
		\centering
		\includegraphics[width=\textwidth]{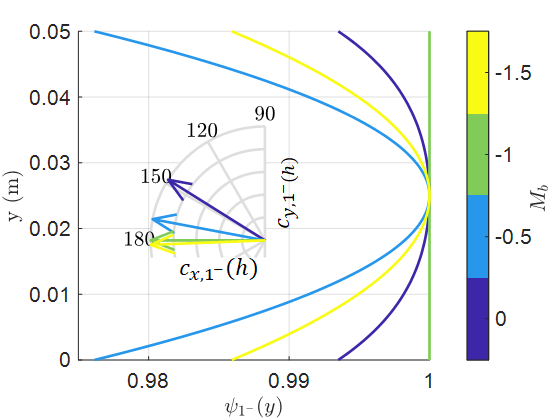}
		\caption{}
		\label{fig:mode1minus_500Hz_varyingNegMb_withCompass}
	\end{subfigure}
	\caption{Shapes of modes $1^+$ \textbf{(a)} and $1^-$ \textbf{(b)} at 500 Hz, and corresponding polar plots of the group velocities normalized to 1, for $\eta_{Loc}=1$ and different $M_b<0$.}
	\label{fig:mode1plusminus_500Hz_varyingMb}
\end{figure}

\begin{figure}[ht!]
	\centering
\includegraphics[width=0.9\textwidth]{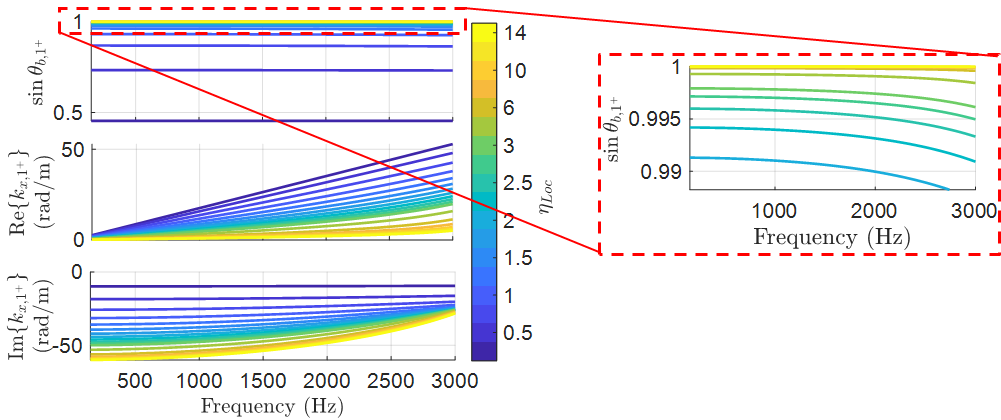}
	\caption{Spectra of $\sin\theta_{b,m}$, $\mathrm{Re}\{k_{x,m}\}$ and $\mathrm{Im}\{k_{x,m}\}$, with varying $\eta_{Loc}$ and $M_b=-1$, for mode $m=1^+$. On the right the zoom of $\sin\theta_{b,1^+}$ close to 1.}
\label{fig:cgnANDReImKxVSfreq_mode9_500Hz_varyingEtaLoc_-1Mb_withZOOM}
\end{figure}

Fig. \ref{fig:mode1plus_500Hz_varyingNegMb_withCompass} shows the variation of the mode $1^+$ shapes for various ABL Mach numbers $M_b<0$, at 500 Hz. Looking at the mode-shapes $\psi_{1^+}(y)$, it is evident how the wavelength along $y$ is significantly reduced for higher absolute values of $M_b<0$. This means a higher normal derivative at the boundary, hence an increase in the modal group velocity along $y$. The normalized vectors $\vec{c}_{1^+}$ at the boundary $y_b=h$ are also reported in Fig. \ref{fig:mode1plus_500Hz_varyingNegMb_withCompass} to illustrate their rotation with $M_b$ varying. For higher absolute values of $M_b<0$, the modal group velocity at the boundary rotates towards the normal to the boundary itself. Fig. \ref{fig:mode1minus_500Hz_varyingNegMb_withCompass} shows the mode $1^-$ shapes for various ABL Mach numbers $M_b<0$, at 500 Hz. Observe that for $M_b=-1$, $\psi_{1^-}=1$ is a plane wave, with group velocity at the boundary directed toward $-x$. Notice also that for $M_b=-1.5$, $c_{1^-}(h)$ has a slightly negative component along $y$. Indeed, for $M_b<-1$, it is $\sin\theta_{b,1^-}<0$ and the propagation of mode $1^-$ is unstable.\\
Now, we want to check the effect of $\eta_{Loc}=1/\zeta_{Loc}$ on the attenuation performances and passivity limits. In Fig. \ref{fig:cgnANDReImKxVSfreq_mode9_500Hz_varyingEtaLoc_-1Mb_withZOOM}, the spectra of $\sin\theta_{b,m}$, $\mathrm{Re}\{k_{x,m}\}$ and $\mathrm{Im}\{k_{x,m}\}$, with varying $\eta_{Loc}$ and $M_b=-1$, are plotted for mode $m=1^+$. Observe that both $\sin\theta_{b,1^+}$ and $|\mathrm{Im}\{k_{x,1^+}\}|$ increases with $\eta_{Loc}$, though the tendency with frequency is different especially at high frequencies as already noticed before. Moreover, $\eta_{Loc}$ does not affect the stability of mode $1^+$ for $M_b=-1$. Fig. \ref{fig:cgnANDReImKxVSfreq_mode9_500Hz_varyingEtaLoc_-1Mb_withZOOM} shows the same plots but relative to mode $1^-$. Observe how $\eta_{Loc}$ has no impact on such mode in case of $M_b=-1$. Indeed, for $M_b=-1$, mode $1^-$ is a plane wave independently of the value assumed by $\eta_{Loc}$.

\begin{figure}[ht!]
	\centering
	\includegraphics[width=0.5\textwidth]{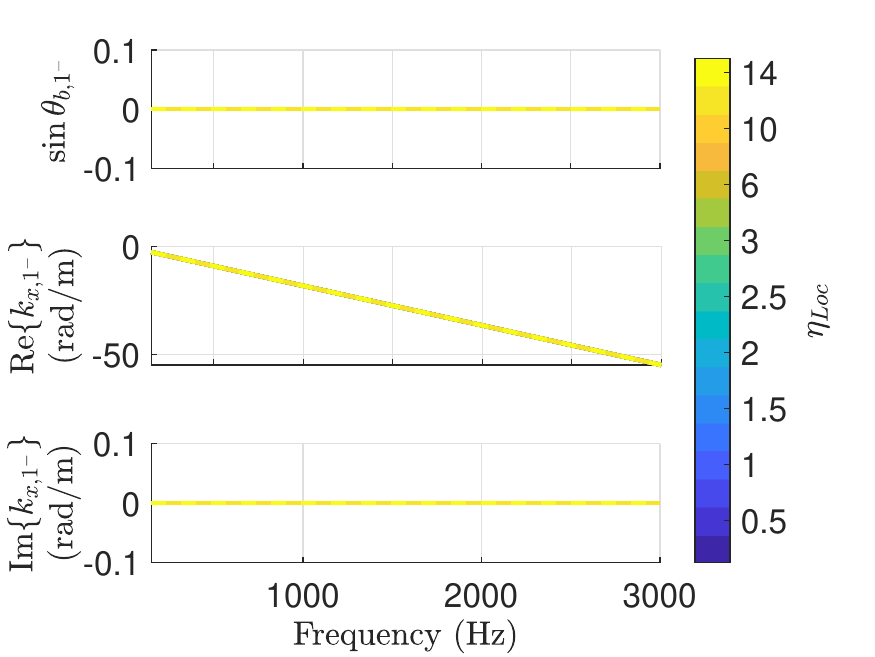}
\caption{Spectra of $\sin\theta_{b,m}$, $\mathrm{Re}\{k_{x,m}\}$ and $\mathrm{Im}\{k_{x,m}\}$, with varying $\eta_{Loc}$ and $M_b=-1$, for mode $m=1^-$}
	\label{fig:cgnANDReImKxVSfreq_mode1_500Hz_varyingEtaLoc_-1Mb}
\end{figure}

\begin{figure}[ht!]
	\centering
	\includegraphics[width=0.9\textwidth]{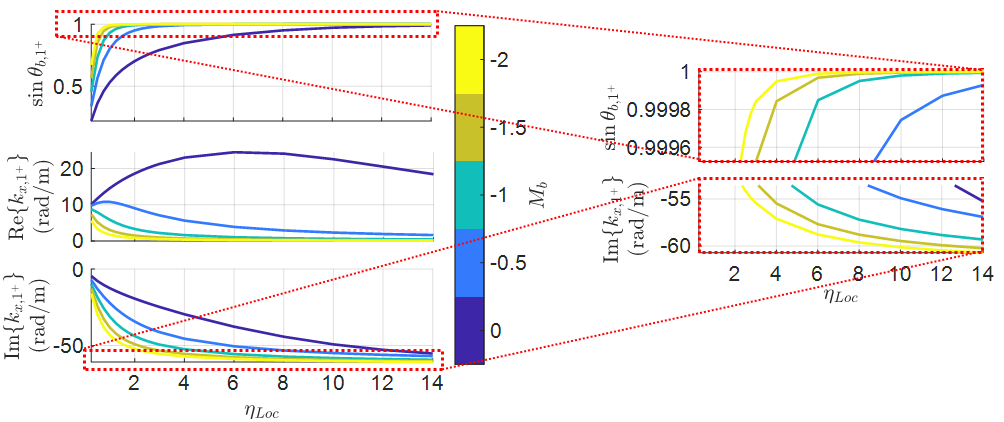}
	\caption{Plots of $\sin\theta_{b,m}$, $\mathrm{Re}\{k_{x,m}\}$ and $\mathrm{Im}\{k_{x,m}\}$, versus $\eta_{Loc}$, with varying $M_b<0$, for mode $m=1^+$. On the right the zoom of $\sin\theta_{b,m}$ close to the maximum value, and $\mathrm{Im}\{k_{x,m}\}$ close to the minimum value.}
	\label{fig:cgnANDReImKxVSetaLoc_mode9_500Hz_varyingMb_withZOOM}
\end{figure}

In Fig. \ref{fig:cgnANDReImKxVSetaLoc_mode9_500Hz_varyingMb_withZOOM}, the same quantities are plotted but for a fixed frequency (500 Hz), against $\eta_{Loc}$ and for varying $M_b<0$, for mode $1^+$. Apparently, increasing $\eta_{Loc}$ improves the attenuation level of mode $1^+$, and its stability is preserved. Fig. \ref{fig:cgnANDReImKxVSetaLoc_mode9_500Hz_varyingMb_withZOOM} reports the same plots but for mode $1^-$. Notice the plane wave solution ($k_{x,1^-}=-k_0$) for $M_b=-1$ which is independent from $\eta_{Loc}$. Also notice that the stability of mode $1^-$ is lost when $M_b<-1$, independently of $\eta_{Loc}$, as long as $\eta_{Loc}$ is purely real and positive. This result confirms the passivity limits found in open field (see Eq. \eqref{eq:passivity condition semi-inf domain}). Hence, we can finally affirm that the ABL passivity limits in open-field (see Section \ref{sec:passivity discussion semi-inf domain}) coincides with the absolute passivity limits of the ABL in the guided grazing incidence problem, in case of purely real $\eta_{Loc}$.

\begin{figure}[ht!]
	\centering
	\includegraphics[width=0.9\textwidth]{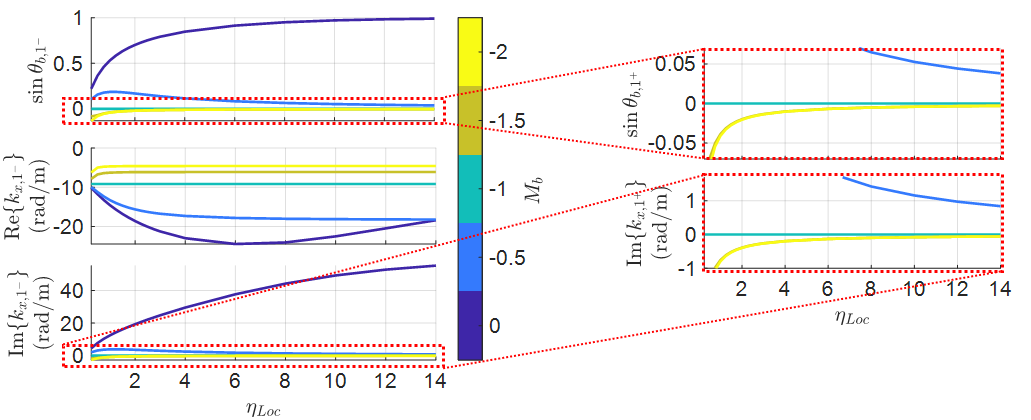}
	\caption{Plots of $\sin\theta_{b,m}$, $\mathrm{Re}\{k_{x,m}\}$ and $\mathrm{Im}\{k_{x,m}\}$, versus $\eta_{Loc}$, with varying $M_b<0$, for mode $m=1^-$. On the right the zoom of $\sin\theta_{b,m}$ and $\mathrm{Im}\{k_{x,m}\}$ close 0.}
	\label{fig:cgnANDReImKxVSetaLoc_mode1_500Hz_varyingMb_withZOOM}
\end{figure}

\begin{figure}[ht!]
	\centering
	\begin{subfigure}[ht!]{0.45\textwidth}
		\centering
		\includegraphics[width=\textwidth]{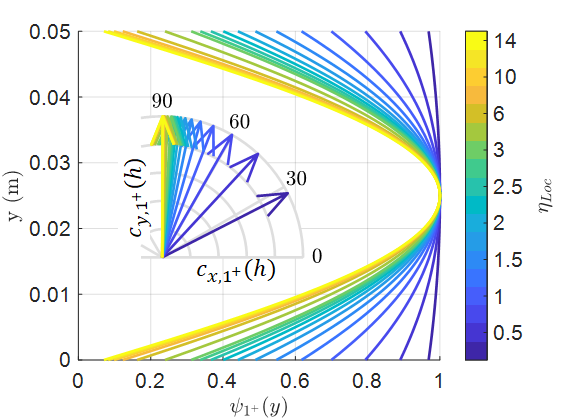}
		\caption{}
		\label{fig:mode1plus_500Hz_varyingEtaLoc_-1Mb_withCompass}
	\end{subfigure}
	\hspace{0.5 cm}
	\begin{subfigure}[ht!]{0.45\textwidth}
		\centering
		\includegraphics[width=\textwidth]{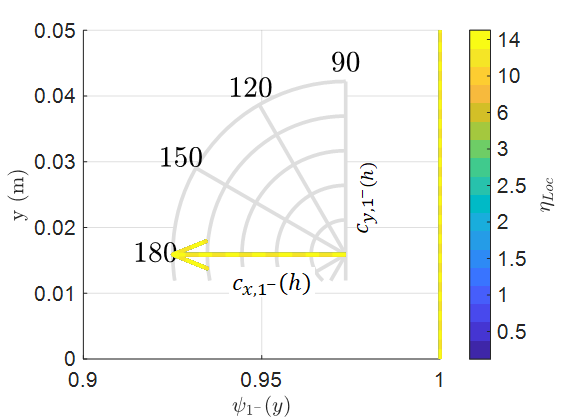}
		\caption{}
		\label{fig:mode1minus_500Hz_varyingEtaLoc_-1Mb_withCompass}
	\end{subfigure}
	\caption{Shapes of modes $1^+$ \textbf{(a)} and $1^-$ \textbf{(b)} at 500 Hz, and corresponding polar plots of the group velocities normalized to 1, for $M_b=-1$ and different $\eta_{Loc}$.}
	\label{fig:mode1plusminus_500Hz_varyingEtaLoc}
\end{figure}

In Fig. \ref{fig:mode1plus_500Hz_varyingEtaLoc_-1Mb_withCompass}, we report the mode $1^+$ shapes for $M_b=-1$ and varying $\eta_{Loc}$, along with the modal group velocity at the boundary. These plots help to visualize the effect of increasing $\eta_{Loc}$, which is very similar to the increase of the absolute value of $M_b<0$, as long as mode $1^+$ is concerned. Fig. \ref{fig:mode1plus_500Hz_varyingEtaLoc_-1Mb_withCompass} confirms that mode $1^-$ stays a plane wave independently of $\eta_{Loc}$, as long as $M_b=-1$.\\

From the duct mode analysis presented in this section, in case of purely real $\eta_{Loc}$, we can affirm that the ABL absolute passivity limits coincide with the passivity limits in open field. The physical quantity $\sin\theta_{b,m}$, other than allowing to define a \emph{modal} passivity criteria, very well correlates with the attenuation rates for mode $1^+$, and could hence be employed for optimization purposes. Moreover, it can help in the interpretation of the physical mechanism behind the enhancement of the attenuation rate achieved by the ABL with respect to purely local impedances.
The physical explanation of the influence of $M_b$ upon the modal propagation angle at the boundary results to be quite intuitive, if compared to the instance of natural convection induced by air-flow blowing into a duct. In that case, waves are naturally convected downstream, with the modal propagation angle increasing for upstream propagating modes \cite{rice1979modal}. This phenomenon explains why, when the duct boundaries are treated by (reciprocal) acoustic liners, and in presence of air-flow, the upstream propagation is more attenuated with respect to the downstream one. In our case, there is no air-flow blowing in the duct. Nevertheless, we can induce an increase of the propagation angle of mode $1^+$ at the boundary, for a fixed $\eta_{Loc}$, by introducing an artificial boundary convection against the propagation of mode $1^+$. This is what an ABL does with $M_b<0$. \\


\subsection{Complex local impedance $\zeta_{Loc}$}\label{sec:duct modes complex impedance}

In this Section, the local impedance component of the ABL is taken as a SDOF resonator, which is the case for most of the actual tunable liners, as the ERs. The mass and stiffness terms of $\zeta_{Loc}$ are taken proportional to the acoustic mass and stiffness of the open-circuit ER prototype employed in the experimental test-bench of Section \ref{sec:experimental grazing incidence}, while the resistance term is taken as a fraction of the characteristic air impedance $\rho_0c_0$. This convention follows the one provided in \cite{de2022effect}. Hence:

\begin{equation}\label{eq:zeta_Loc complex}
	\zeta_{Loc}(\mathrm{j}\omega)=\frac{1}{\rho_0c_0} \biggr(M_{d}\mathrm{j}\omega + R_{d} + \frac{K_{d}}{\mathrm{j}\omega}\biggr),
\end{equation}

where $R_d=r_d\rho_0c_0$ is the desired resistance, while the desired reactive components are defined as $M_{d}=\mu_M M_{0}$ and $K_{d}=\mu_K K_{0}$, with $M_0$ and $K_0$ the acoustic mass and stiffness of the open-circuit ER prototype employed in the experimental test-bench of Section \ref{sec:experimental grazing incidence}. Their values are reported in Table \ref{tab:TSparam FEMTO}. The resonance frequency of $\zeta_{Loc}$ can be varied by tuning either the stiffness $\mu_K$ or the mass $\mu_M$ parameters, as $f_d=f_0\sqrt{\mu_K/\mu_M}$, with $f_0$ being the resonance frequency of the open-circuit ER (468 Hz). Reducing $\mu_M=\mu_K$, or increasing $r_d$, allows to reduce the quality factor of the SDOF resonator.

\begin{table}
	\centering
	\begin{tabular}{c|cccc}
		\hline
		\hline
		Model parameters & $M_{0}$ & $R_{0}$ & $K_{0}$ & $Bl/S_e$\\
		\hline
		Units & $\mathrm{kg/m^2}$ & $\mathrm{Pa.s/m}$ & $\mathrm{Pa/m}$ & $\mathrm{Pa.A^{-1}}$ \\
		\hline
		Values & $0.342$ & $133$ & $2.96\times10^{6}$ & $846$ \\
		\hline
		\hline
	\end{tabular}
	\caption{Model parameters of the ER. The values of $R_0$ and $Bl/S_e$ are provided for results shown in Sections \ref{sec:scatt 3D} and \ref{sec:experimental grazing incidence}.}
	\label{tab:TSparam FEMTO}
\end{table} 


\begin{figure}[ht!]
		\centering
		\includegraphics[width=0.7\textwidth]{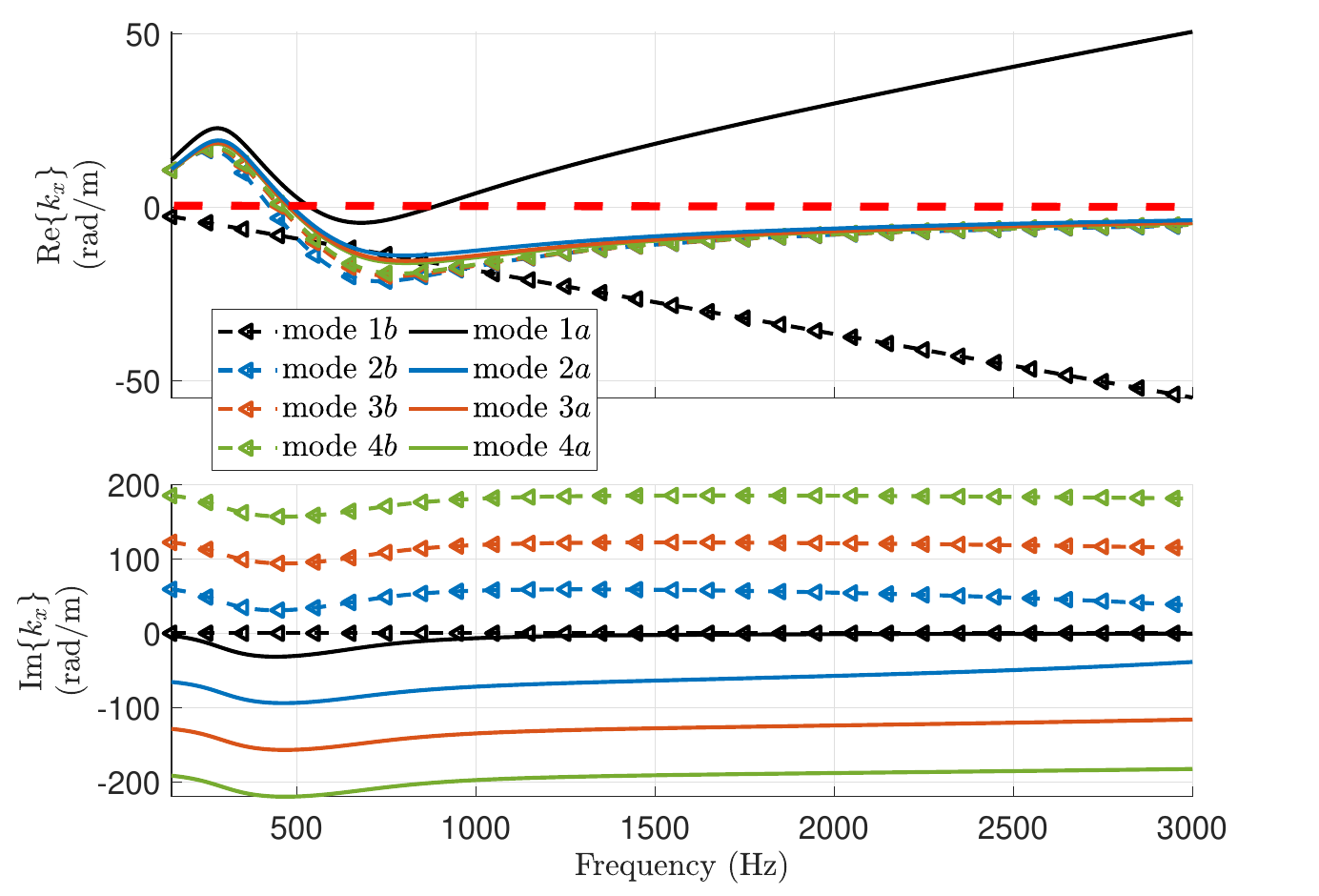}
		\centering
	\caption{Dispersion plots relative to the first eight duct modes in case of ABL with complex $\zeta_{Loc}(\mathrm{j}\omega)$ given by Eq. \eqref{eq:zeta_Loc complex}. The control parameters are set to $\mu_M=\mu_K=0.5$, $r_d=1$, and $M_b=-1$. In dashed red is the line $\mathrm{Re}\{k_x\}=0$}
		\label{fig:K_Omega_plane_NonLocalBC_RatRho0c0_Kat0.2_Mat0.2_-1cac0}
\end{figure}

Fig. \ref{fig:K_Omega_plane_NonLocalBC_RatRho0c0_Kat0.2_Mat0.2_-1cac0} shows the dispersion plots of $k_x(\omega)$ for $\mu_M=\mu_K=0.5$, $r_d=1$, and $M_b=-1$. The modes are not labelled referring to their sense of propagation (positive or negative) as the Re$\{k_x\}$ happens to change its sign with frequency, while sign of Im$\{k_x\}$ is unaltered. Mode $1b$ corresponds to the backward propagating plane wave always present for $M_b=-1$. Mode $1a$ is the first mode propagating toward $+x$. Nevertheless, Re$\{k_{x,1a}\}$ becomes negative between approximately 500 and 870 Hz, which means a reverse in the direction of propagation. In such frequency range, Re$\{k_x\}$ and Im$\{k_x\}$ present the same sign, which means unstable propagation. Therefore, we can state that, for $\mu_M=\mu_K=0.5$, $r_d=1$ and $M_b=-1$, the ABL does not fulfil the modal passivity criteria for any mode, except for mode $1b$. Hence, the absolute passivity conditions Re$\{\zeta_{Loc}\}\geq0$ and $|M_b|<1$, sufficient in case of open-field or purely resistive $\zeta_{Loc}$, will not be sufficient in case of reactive $\zeta_{Loc}$. 
As this phenomenon does not happen for purely real $\zeta_{Loc}$ (see Section \ref{sec:duct modes real impedance}), neither in case of locally-reacting boundary ($M_b=0$), we expect that the ABL could restore modal passivity by reducing the reactive character of $\zeta_{Loc}$, or by decreasing $M_b$. It is interesting to highlight that such non-passive behaviour cannot be detected in open-field, suggesting also the influence of the cross-section dimension in the duct-modes stability.

\begin{figure}
	\centering
	\begin{subfigure}[ht!]{0.45\textwidth}
		\centering
		\includegraphics[width=\textwidth]{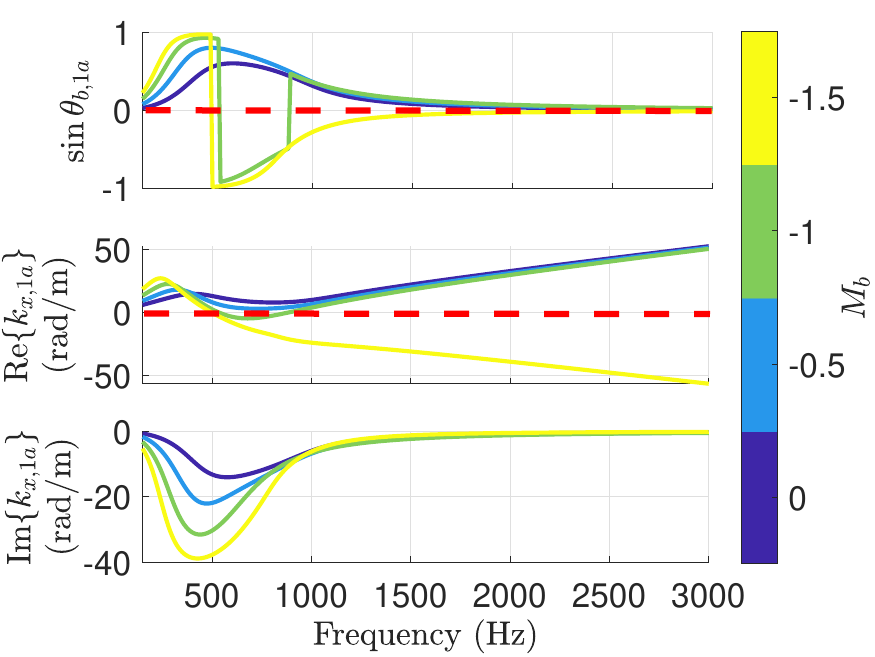}\\
		\caption{}
		\label{fig:cgnANDReImKxVSfreq_mode9_varying_Mb}
	\end{subfigure}	
	\hspace{1 cm}
	\begin{subfigure}[ht!]{0.45\textwidth}
		\centering
		\includegraphics[width=\textwidth]{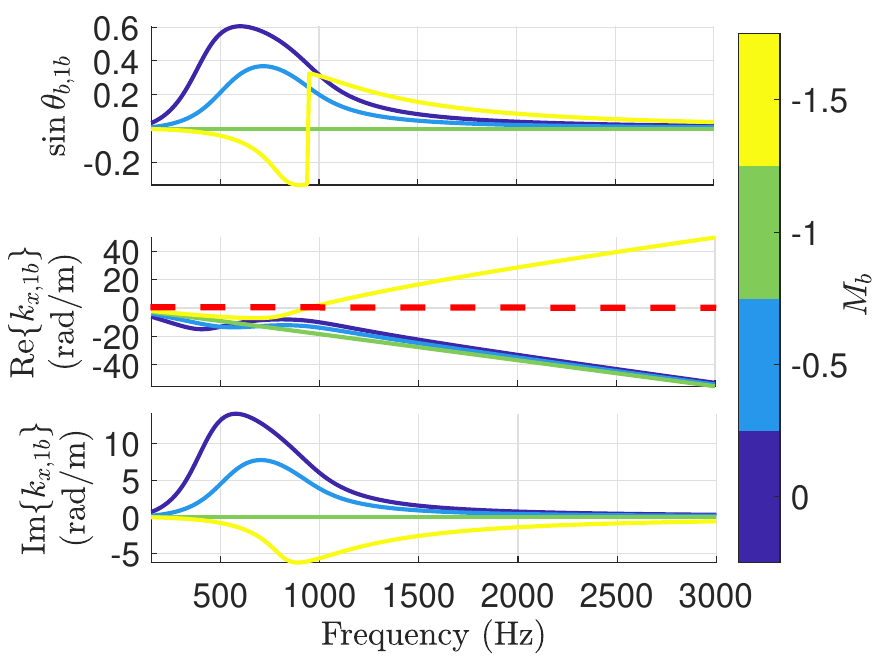}\\
		\caption{}
		\label{fig:cgnANDReImKxVSfreq_mode1_varying_Mb}
	\end{subfigure}
	\caption{Spectra of $\sin\theta_{b,m}$, real and imaginary parts of $k_{x,m}$, for mode $1a$ \textbf{(a)} or $1b$ \textbf{(b)}, with $M_b$ varying. The other parameters are set to $\mu_M=\mu_K=0.5$, $r_d=1$, and the duct cross section width is $h=0.05$ m. The dashed red lines separate passive and non passive conditions.}
	\label{fig:cgnANDReImKxVSfreq_mode9and1_varying_Mb}
\end{figure}

\begin{figure}
	\centering
	\begin{subfigure}[ht!]{0.45\textwidth}
		\centering
		\includegraphics[width=\textwidth]{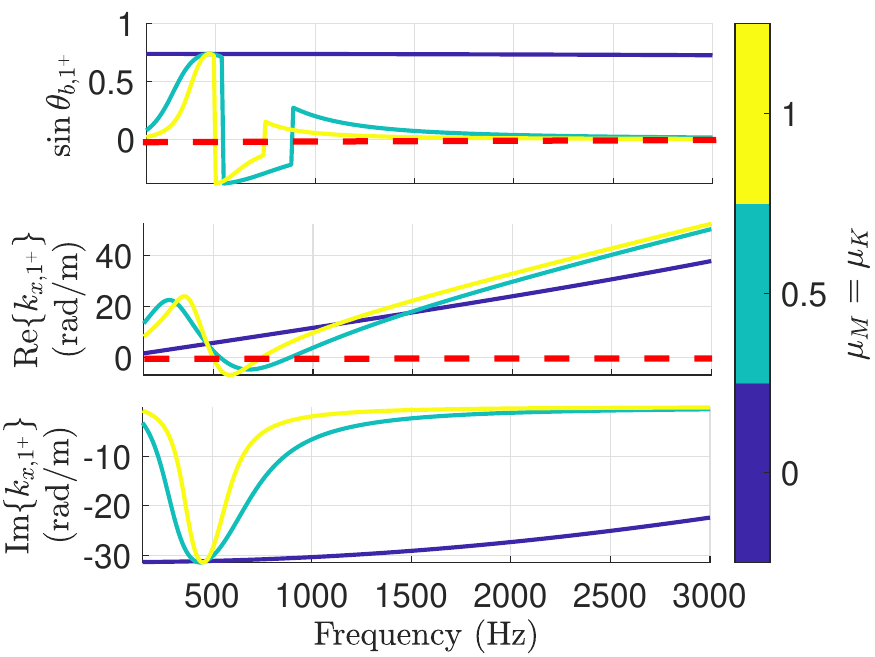}\\
		\caption{}
		\label{fig:cgnANDReImKxVSfreq_mode9_varying_muMmuK}
	\end{subfigure}	
	\hspace{1 cm}
	\begin{subfigure}[ht!]{0.45\textwidth}
		\centering
		\includegraphics[width=\textwidth]{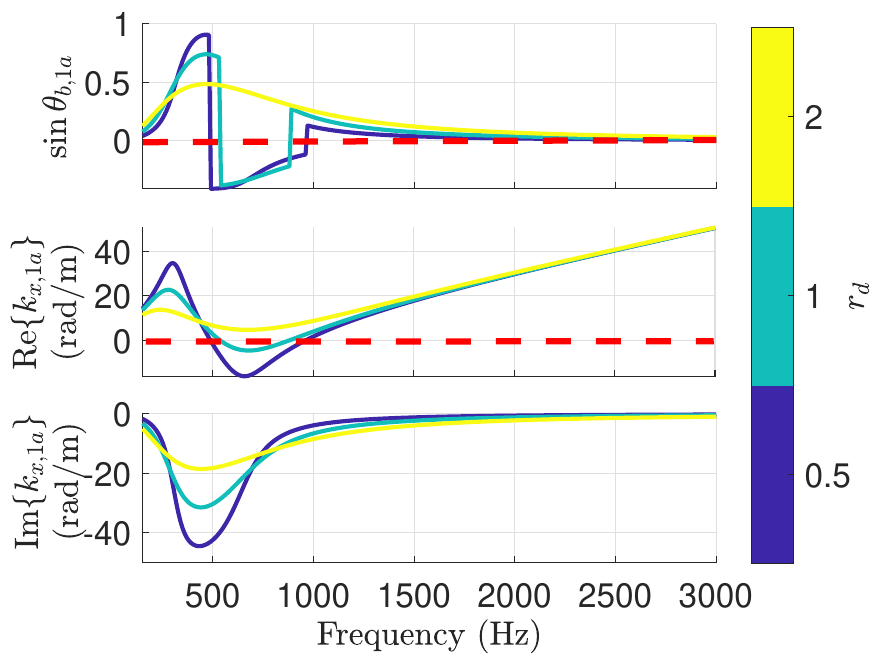}\\
		\caption{}
		\label{fig:cgnANDReImKxVSfreq_mode9_varying_rd}
	\end{subfigure}
	\caption{Spectra of $\sin\theta_{b,m}$, real and imaginary parts of $k_{x,m}$, for mode $1a$, with varying $\mu_M=\mu_K$ \textbf{(a)} or $r_d$ \textbf{(b)}. The default parameters are set to $\mu_M=\mu_K=0.5$, $r_d=1$, $M_b=-1$, and the duct cross section width is $h=0.05$ m. The dashed red lines separate passive and non passive conditions.}
	\label{fig:cgnANDReImKxVSfreq_mode9_varying_QualityFactor}
\end{figure}

Fig. \ref{fig:cgnANDReImKxVSfreq_mode9_varying_Mb} shows the spectra of $\sin\theta_{b,1a}$ and of the real and imaginary parts of $k_{x,1^+}$, with varying $M_b<0$. Increasing $|M_b|$ leads to higher values of Im$\{k_{x,1a}\}$ around resonance (as in case of purely resistive $\zeta_{Loc}$). Though, for $M_b=-1$ and $-1.5$, the Re$\{k_{x,1a}\}$ changes its sign in a frequency range starting just above resonance. The higher $|M_b|$, the larger is such frequency band of non-passive behaviour. Indeed, while for $M_b=-1$ passivity is restored at about 900 Hz, in case of $M_b=-1.5$, passivity is never restored in the frequency range under study. It is remarkable the correlation of $\sin\theta_{b,1a}$ with both the acoustical passivity and the attenuation rate. Indeed, the frequency bandwidth where Re$\{k_{x,1a}\}$ changes its sign, coincides perfectly with the bandwidth where $\sin\theta_{b,1a}<0$. Moreover, in the passivity regions, $\sin\theta_{b,1a}$ is higher when Im$\{k_{x,1a}\}$ presents larger values, thus confirming the correlation with the attenuation rate. In Fig. \ref{fig:cgnANDReImKxVSfreq_mode1_varying_Mb}, the same modal quantities are plotted with varying $M_b<0$, but for mode $1b$. Notice the plane wave solution for $M_b=-1$. For $M_b=-1.5$, the Im$\{k_{x,1b}\}$ becomes negative as the Im$\{k_{x,1^-}\}$ in case of purely real $\zeta_{Loc}$ (check Fig. \ref{fig:cgnANDReImKxVSfreq_mode1_varyingMb_PurelyRealEtaLoc}). Nevertheless, the Re$\{k_{x,1^+}\}$ changes its sign, therefore restraining the non-passive behaviour up to about 950 Hz. After this frequency, passivity gets restored. Once again, check the perfect correlation of the dispersion solutions with the values of $\sin\theta_{b,1b}$, both in terms of passive bandwidth and attenuation rates. Fig. \ref{fig:cgnANDReImKxVSfreq_mode9_varying_QualityFactor} shows the effect of the quality factor of $\zeta_{Loc}$ upon the modal quantities of mode $1a$, for $M_b=-1$. In particular, Fig. \ref{fig:cgnANDReImKxVSfreq_mode9_varying_muMmuK} shows the effect of the reactive terms $\mu_M=\mu_K$, while Fig. \ref{fig:cgnANDReImKxVSfreq_mode9_varying_rd} shows the effect of the resistive one $r_d$. As expected, by reducing the quality factor of $\zeta_{Loc}$ (by decreasing $\mu_M=\mu_K$ and/or augmenting $r_d$), we can restore the acoustical passivity. Once again, both passivity limits and attenuation rates are perfectly captured by $\sin\theta_{b,1a}$.

\begin{figure}[ht!]
	\centering
	\includegraphics[width=0.5\textwidth]{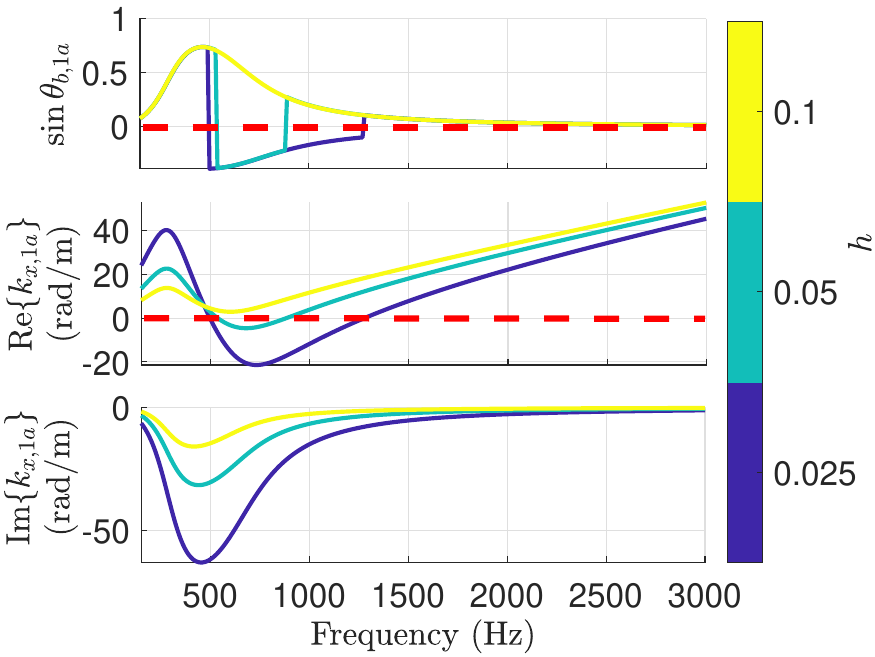}
	\centering
	\caption{Spectra of $\sin\theta_{b,m}$, real and imaginary parts of $k_{x,m}$, for mode $1a$, with varying $h$. The control parameters are set to $\mu_M=\mu_K=0.5$, $r_d=1$ and $M_b=-1$. The dashed red lines separate passive and non passive conditions.}
	\label{fig:cgnANDReImKxVSfreq_mode9_varying_h}
\end{figure}

Finally, we want to check the effect of the duct cross-section width $h$. In Fig. \ref{fig:cgnANDReImKxVSfreq_mode9_varying_h}, $h$ is halved and doubled with respect to the default value, demonstrating that such non-passive behaviour is strictly related to the duct cross-section size. The narrower the duct cross-section is, the larger is the bandwidth of passivity loss.
It is also interesting to remark that $\sin\theta_{b,1a}$, despite perfectly capturing the frequency ranges of non-passive behaviour, is not able to capture the variation of attenuation level (Im$\{k_{x,1a}\}$) with $h$. 
It looks like the boundary modal group velocity (which gives $\sin\theta_{b,1a}$) is not informed by the variation of the duct cross section size, except if $h$ leads to a change of direction of propagation. A deeper analysis of the modal group velocity $\vec{c}_{m}$ on the boundary, and the effect of $h$ upon it, is out of the scope of the present paper, but it will be retrieved in a future study. Nevertheless, as $h$ is not a parameter related to the boundary operator, the quantity $\sin\theta_{b,m}$ can still be employed for liner optimization purposes.\\

Below, the main outcomes of Section \ref{sec:duct modes analysis}:

\begin{itemize}
	\item \textbf{In case of purely real $\zeta_{Loc}$, the ABL absolute passivity limits in duct grazing incidence, coincide with the passivity limits in open field.}
	\item \textbf{For $|M_b|>1$, modal passivity is lost for the first mode propagating in the same direction as $M_b$, while it is preserved for duct modes propagating against $M_b$.}
	\item \textbf{The ABL boundary convection increases $\sin\theta_{b,m}$ for modes propagating against $M_b$, analogously to air-flow convection which increases the modal propagation angle of upstream propagating modes  \cite{rice1979modal}.}
	\item \textbf{In case of complex $\zeta_{Loc}$, the open field passivity criteria do not coincide with the absolute passivity conditions in grazing incidence.}
	\item \textbf{In case of complex $\zeta_{Loc}$, mode $1a$ passivity is reduced when: the quality factor of $\zeta_{Loc}$ increases, $|M_b|$ augments, and the duct cross-section width is reduced.}
	\item \textbf{$\sin\theta_{b,m}$ provides a criteria for both passivity and attenuation rates, both in case of purely real and complex $\zeta_{Loc}$.}
	\item \textbf{Non-reciprocal propagation is evident by the first duct-mode solutions: mode $1^+$ (read $1a$ for complex $\zeta_{Loc}$) is attenuated, while mode $1^-$ (read $1b$ for complex $\zeta_{Loc}$) is a plane wave, for $M_b=-1$ and independently from $\zeta_{Loc}$.}
\end{itemize}

\FloatBarrier

\section{Scattering simulations in 2D waveguide}\label{sec:scattering 2D}

\begin{figure}[ht!]
	\centering
	\begin{subfigure}[ht!]{0.45\textwidth}
		\centering
		\includegraphics[width=\textwidth]{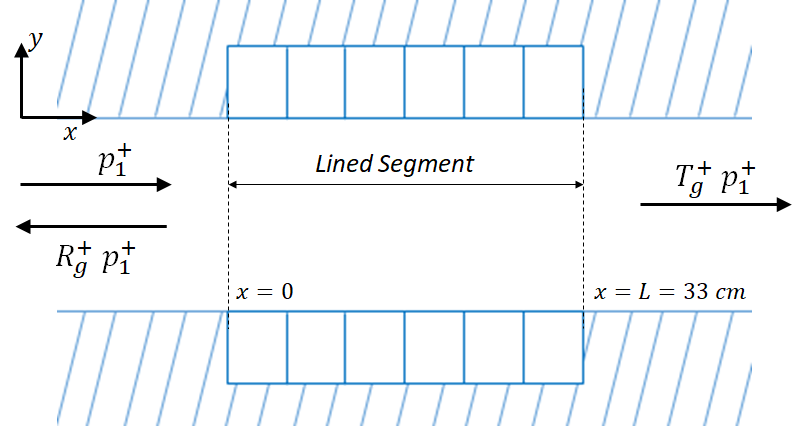}
		\centering
		\caption{}
		\label{fig:scattering_coeff_def_plus}
	\end{subfigure}
	\hspace{1 cm}
	\begin{subfigure}[ht!]{0.45\textwidth}
		\centering
		\includegraphics[width=\textwidth]{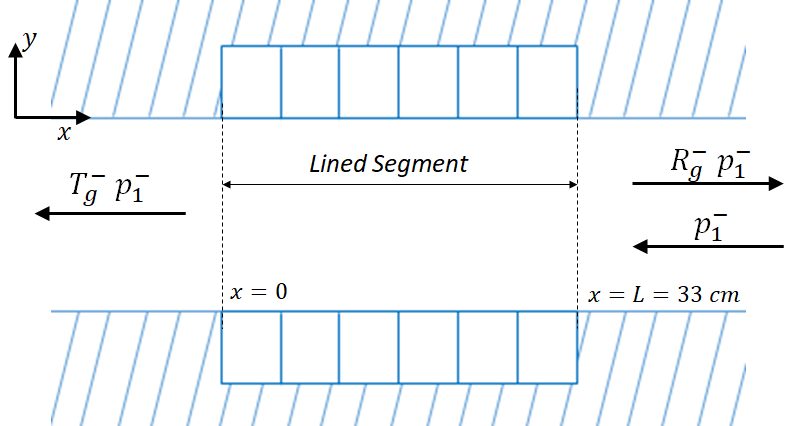}
		\centering
		\caption{}
		\label{fig:scattering_coeff_def_minus}
	\end{subfigure}
	\caption{Lining segment and scattering coefficients definition in a 2D waveguide lined on both sides.}
	\label{fig:scattering_coeff_def_plusminus}
\end{figure}

In this section the ABL is analysed in terms of scattering performances in the plane wave regime. The liner is considered to extend for an axial length $L=0.3$ m in a 2D acoustic waveguide of cross-section height $h=0.05$ m, without flow. Such dimensions correspond to the experimental setup that will be presented in Section \ref{sec:experimental grazing incidence}. The scattering problem is illustrated in Fig. \ref{fig:scattering_coeff_def_plusminus}, where the reflection $R_g$ and transmission $T_g$ coefficients are defined for incident field directed toward either $+x$ or $-x$. The subscript $g$ is employed to differentiate the present \emph{grazing} incidence from the oblique incidence scattering of Section \ref{sec:passivity discussion semi-inf domain}. The ABL is applied continuously on the boundary of the waveguide in the lined segment. The scattering matrix is defined in Eq. \eqref{eq:scattering matrix definition} for the plane wave regime of a hard-walled duct.

\begin{equation}\label{eq:scattering matrix definition}
	\begin{bmatrix}
		p_2^+\\p_1^-
	\end{bmatrix}
	=\begin{bmatrix}
		T_g^+ & R_g^- \\
		R_g^+ & T_g^-
	\end{bmatrix}
	\begin{bmatrix}
		p_1^+\\p_2^-
	\end{bmatrix}.
\end{equation}

The superscript signs $+$ or $-$ in Eq. \eqref{eq:scattering matrix definition}, indicate the direction of propagation of the incident plane wave (toward either $+x$ or $-x$).
The results in terms of scattering matrix coefficients, have been obtained by FE simulations in Comsol. As in the duct mode analysis, the FE mesh has been built sufficiently fine to fully resolve both longitudinal and transversal pressure field up to $f_{max}=3$ kHz.
The scattering coefficients $T_g^\pm$ and $R_g^\pm$ are computed, by exciting first the left and then the right termination. In the scattering problem, high noise isolation toward $+x$ ($-x$) corresponds to low values of $T_g^+$ ($T_g^-$). The acoustical passivity, in the plane wave regime, corresponds to positive values of both $\alpha_g^+$ and $\alpha_g^-$.\\
As in the duct mode analysis, we differentiate the case of purely real or resonant $\zeta_{Loc}$ in the ABL.

\subsection{Real local impedance $\zeta_{Loc}$}\label{sec:scattering real impedance}

The scattering performances are presented in terms of power scattering coefficients for both positive and negative propagation. The power scattering coefficients are defined from the power balance \cite{ingard2009noise} which, in case of plane waves, reduces to:

\begin{equation}\label{eq:power balance}
1 = \alpha_g^\pm + |T_g^\pm|^2 + |R_g^\pm|^2,
\end{equation}

where $R_g$ and $\alpha_g$ are the reflection and absorption coefficients in grazing incidence, respectively. From $|T_g^\pm|^2$, it is possible to compute the Transmission Loss $(TL_g^\pm)_{Liner}=10\log_{10}(1/|T_g^\pm|^2)$, and the Insertion Loss $IL_g^\pm=(TL_g^\pm)_{Liner}-(TL^\pm)_{Rigid}$. As $(TL^\pm)_{Rigid}=0$ in simulations, $IL^\pm=(TL_g^\pm)_{Liner}$.

\begin{figure}[ht!]
	\centering
	\begin{subfigure}[ht!]{0.45\textwidth}
		\centering
		\includegraphics[width=\textwidth]{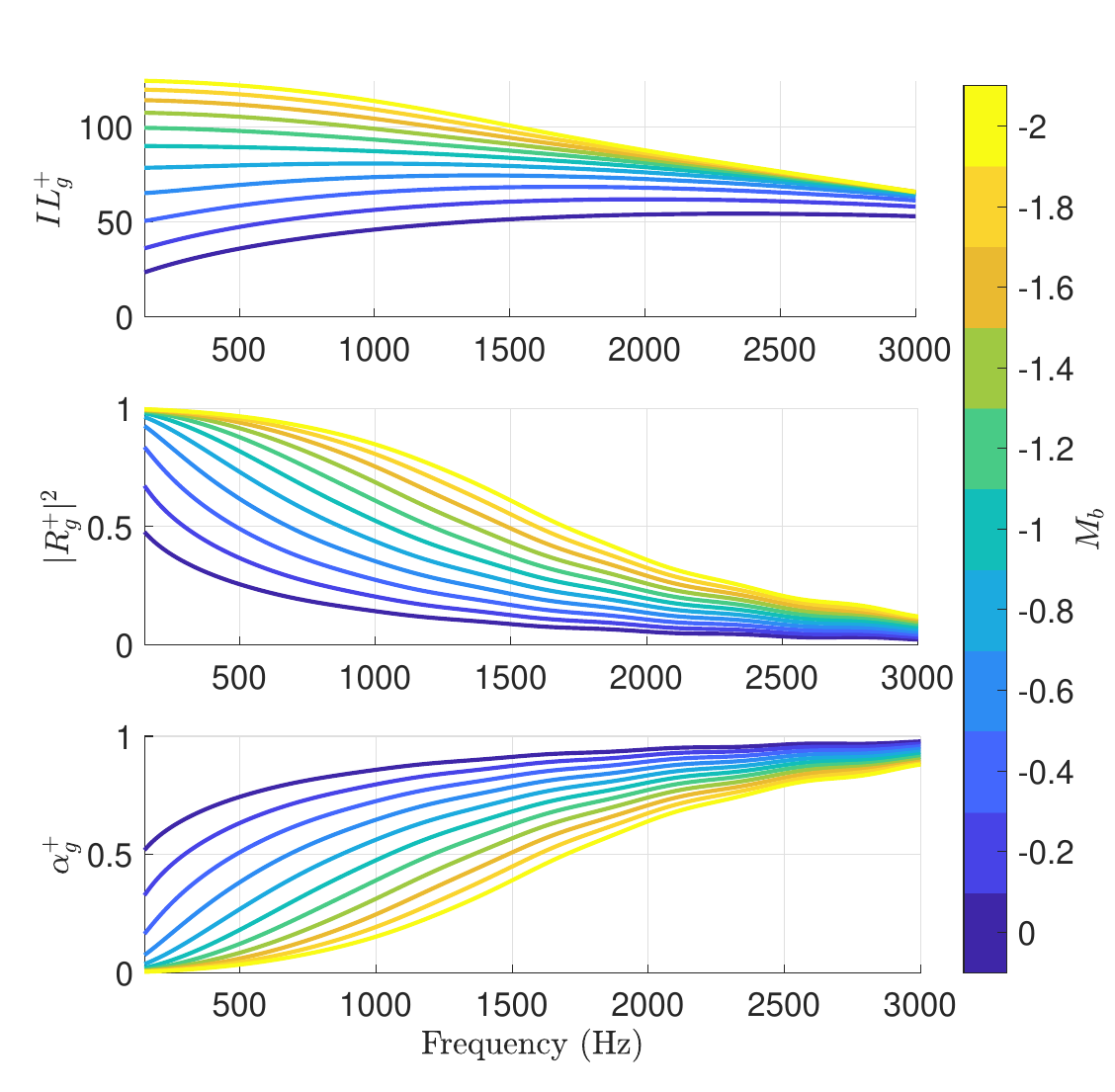}
		\centering
		\caption{}
		\label{fig:ScattCoeffplus_1rho0c0Rat_0muMmuK_0tau_VARYINGcac0}
	\end{subfigure}
	\hspace{1 cm}
	\begin{subfigure}[ht!]{0.45\textwidth}
		\centering
		\includegraphics[width=\textwidth]{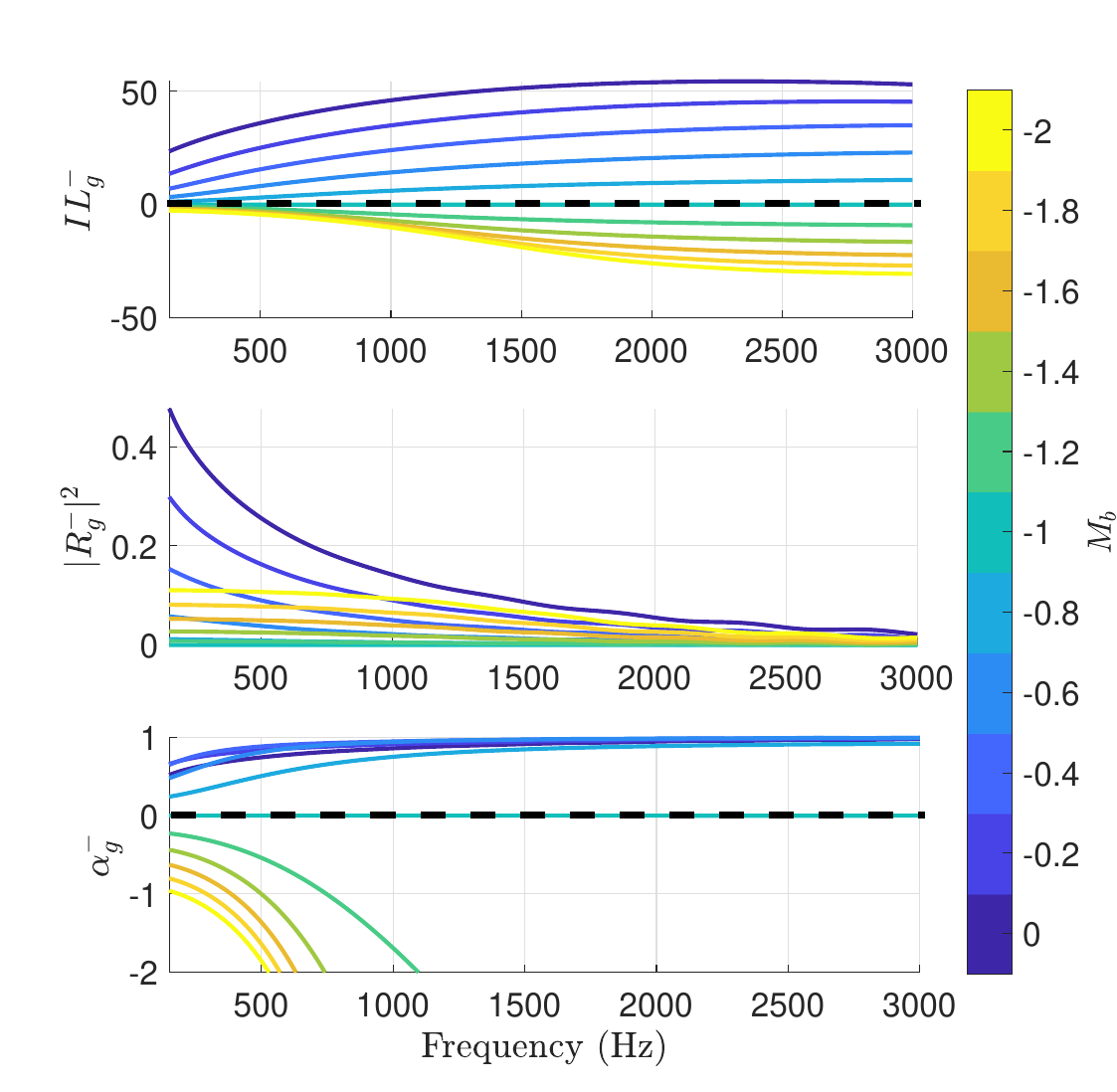}
		\centering
		\caption{}
		\label{fig:ScattCoeffminus_1rho0c0Rat_0muMmuK_0tau_VARYINGcac0}
	\end{subfigure}
	\caption{Scattering coefficients in a 2D waveguide of cross section width $h=0.05$ m with lined segment of length $L=0.3$ m, lined on both sides by the boundary advection law with $\zeta_{Loc}=1$, and varying $M_b$.}
	\label{fig:ScattCoeff_1rho0c0Rat_0muMmuK_0tau_VARYINGcac0}
\end{figure}

Fig. \ref{fig:ScattCoeffplus_1rho0c0Rat_0muMmuK_0tau_VARYINGcac0} shows the power scattering coefficients in case of $\zeta_{Loc}=1$, for $M_b$ continuously varying from $0$ to $-2$. Coherently with the duct mode $1^+$ solution reported in Section \ref{sec:duct modes real impedance}, increasing the absolute value of $M_b<0$, brings about an increase in the $IL_g^+$, especially at low frequencies. Observe that such increase of $IL_g^+$ is accompanied by a significant increment of the back-reflection and, less intuitively, by a reduction of absorption. This means that, in such configuration of waveguide with both upper and lower sides lined by the ABL, excited by plane waves propagating against the boundary advection speed, most energy is reflected back rather being absorbed. In case of negative propagation, i.e. plane waves propagating concordant with $M_b$, perfect transmission is assured for $M_b=-1$, while for $M_b<-1$, the loss of passivity ($\alpha_g^{-1}<0$) of the ABL manifests itself by $|T_g^-|>1$ in agreement with the change of sign of Im$\{k_{x,1^+}\}$ showed in Fig. \ref{fig:cgnANDReImKxVSfreq_mode1_varyingMb_PurelyRealEtaLoc}. The passivity limits are highlighted by dashed black line in Fig. \ref{fig:ScattCoeff_1rho0c0Rat_0muMmuK_0tau_VARYINGcac0}. These results are totally coherent with the results of Section \ref{sec:duct modes real impedance} both in terms of attenuation performances and passivity. Moreover, perfect non-reciprocal propagation is achieved for $M_b=-1$, as $IL_g^-=0$, while $IL_g^+$ is very high. This, also, is in agreement with the dispersion solutions of Section \ref{sec:duct modes real impedance}.

\subsection{Complex local impedance $\zeta_{Loc}$}\label{sec:scatt complex impedance}

\begin{figure}[ht!]
	\centering
	\begin{subfigure}[ht!]{0.45\textwidth}
		\centering
		\includegraphics[width=\textwidth]{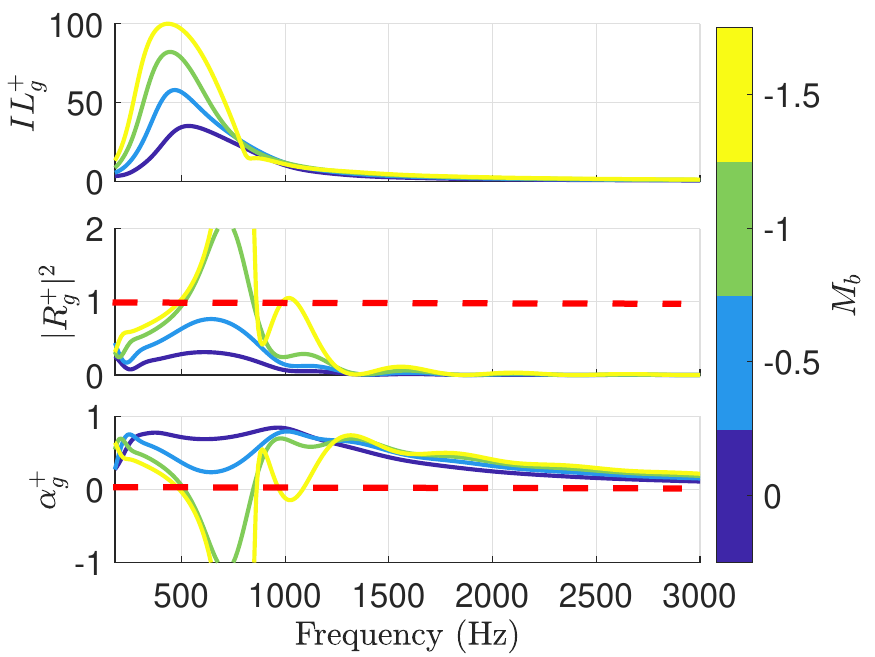}
		\centering
		\caption{}
		\label{fig:ScattPowCoeff_LEFTexc_MbVarying}
	\end{subfigure}
	\hspace{1 cm}
	\begin{subfigure}[ht!]{0.45\textwidth}
		\centering
		\includegraphics[width=\textwidth]{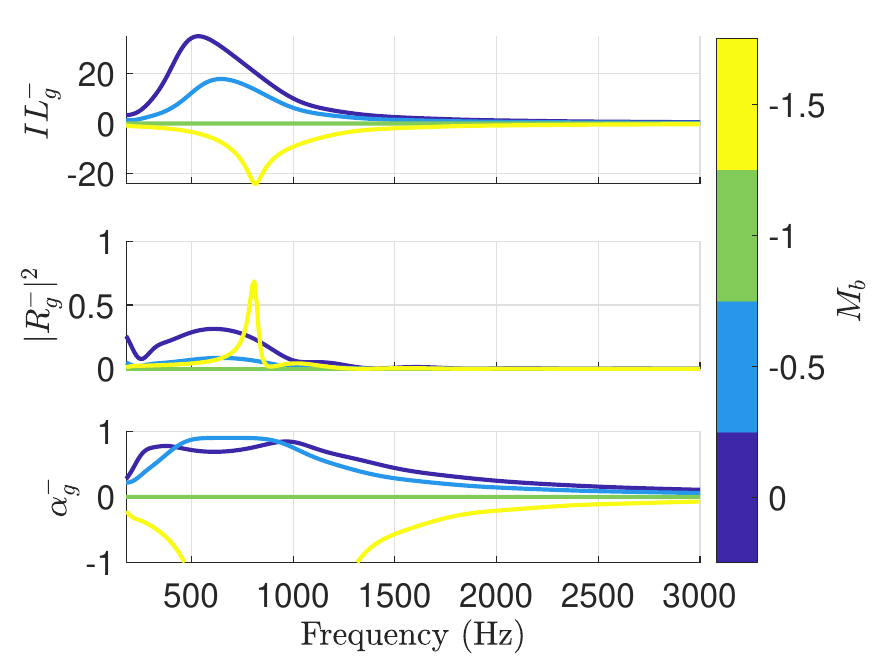}
		\centering
		\caption{}
		\label{fig:ScattPowCoeff_RIGHTexc_MbVarying}
	\end{subfigure}
	\caption{Scattering coefficients for excitation coming from the left \textbf{(a)} or right \textbf{(b)} termination, in a 2D waveguide of cross section height $h=0.05$ m with lined segment length $L=0.3$ m, and ABL applied on both sides of the duct, with $\mu_M=\mu_K=0.5$, $r_{d}=1$ and varying $M_b$.}
	\label{fig:ScattPowCoeff_LEFTandRIGHTexc_MbVarying}
\end{figure}

\begin{figure}[ht!]
	\centering
	\begin{subfigure}[ht!]{0.45\textwidth}
		\centering
		\includegraphics[width=\textwidth]{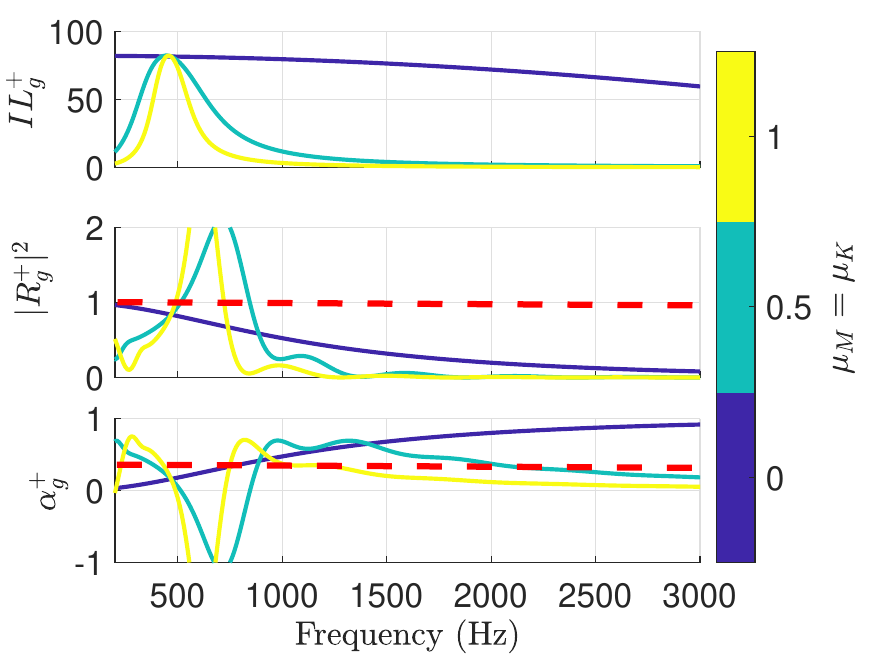}
		\centering
		\caption{}
		\label{fig:ScattPowCoeff_LEFTexc_muMmuKVarying}
	\end{subfigure}
	\hspace{1 cm}
	\begin{subfigure}[ht!]{0.45\textwidth}
		\centering
		\includegraphics[width=\textwidth]{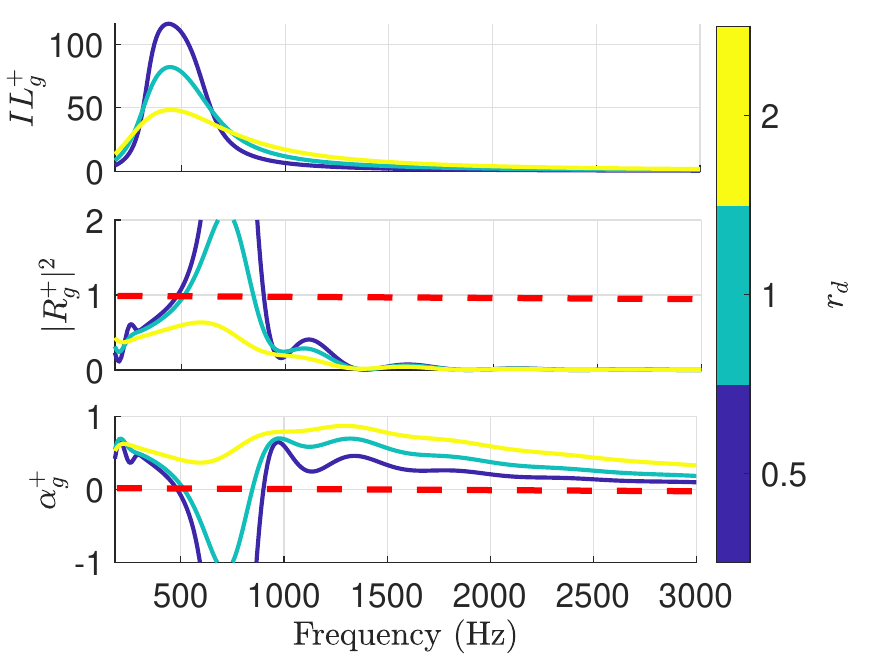}
		\centering
		\caption{}
		\label{fig:ScattPowCoeff_LEFTexc_RdVarying}
	\end{subfigure}
	\caption{Scattering coefficients for excitation coming from the left, in a 2D waveguide of cross section width $h=0.05$ m with an ABL lining both sides for an extension of $L=0.3$ m, in case of $M_b=-1$, and with varying $\mu_M=\mu_K$ \textbf{(a)}, or varying $r_{d}$ \textbf{(b)}. The default values are $\mu_M=\mu_K=0.5$ and $r_{d}=1$.}
	\label{fig:ScattCoeff_LEFTexc_VARYINGmuMmuKRat}
\end{figure}

\begin{figure}[ht!]
		\centering
		\includegraphics[width=0.6\textwidth]{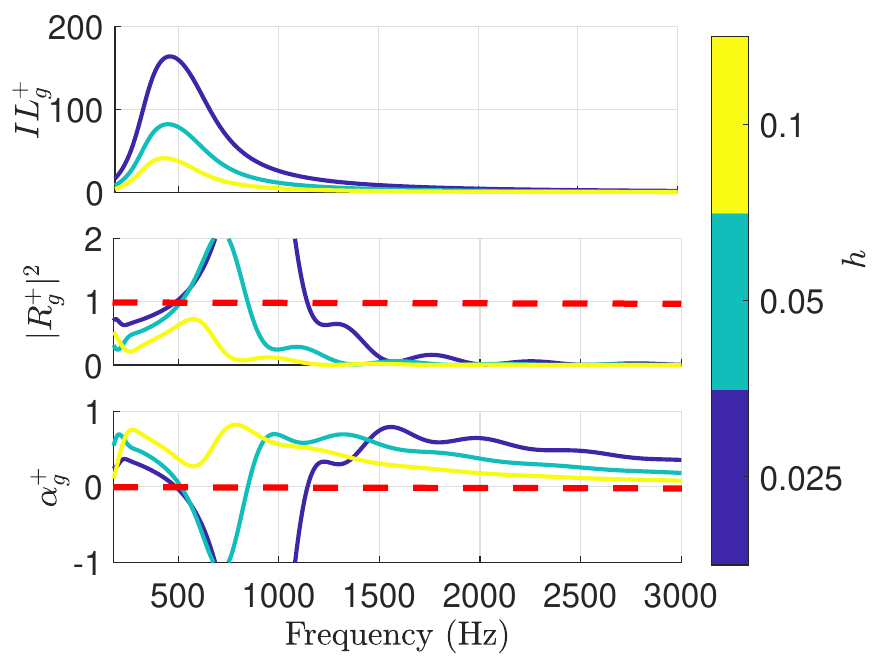}
		\centering
\caption{Scattering coefficients in a 2D waveguide of variable cross section width $h$, lined segment of length $L=0.3$ m, lined on both sides by the ABL with $M_b=-1$, $R_{d}=\rho_0c_0$ and $\mu_M=\mu_K=0.5$.}
\label{fig:ScattPowCoeff_LEFTexc_hVarying}
\end{figure}

As in Section \ref{sec:duct modes complex impedance}, we consider here the scattering problem in case of $\zeta_{Loc}$ assuming the SDOF resonator form of Eq. \eqref{eq:zeta_Loc complex}, with default mass and stiffness coefficients $\mu_M=\mu_K=0.5$, and resistance term $r_d=1$. Fig. \ref{fig:ScattPowCoeff_LEFTexc_MbVarying} shows the effect of varying $M_b$ in case of incoming field from the left duct termination, indicated by $+$ superscript. As expected, increasing the absolute value of $M_b<0$, improves isolation (augments $IL_g^+$). But, after the resonance of $\zeta_{Loc}$, $\alpha_g^+$ becomes negative for $M_b=-1$, up to about 870 Hz. This loss of passivity corresponds to a reflection coefficient higher than 1, in agreement with the change of sign of Re$\{k_{x,1a}\}$ in Fig. \ref{fig:cgnANDReImKxVSfreq_mode9_varying_Mb}, which becomes negative at about 500 Hz, and comes back to be positive at about 870 Hz. Notice the interesting correlation between an unstable propagation of mode $1a$ toward $-x$ in the range $500-870$ Hz, and the reflection coefficient higher than 1 in the same frequency range. Since the duct mode instability manifests as a \emph{backward} propagation, this translates into higher acoustic energy reflected \emph{backward}. Also for $M_b=-1.5$ in Fig. \ref{fig:cgnANDReImKxVSfreq_mode9_varying_Mb}, we have non-passive behaviour, as expected, corresponding once again to a $|R_g^+|>1$. Nevertheless, passivity is restored around $870$ Hz, then lost again in a narrow bandwidth around 1000 Hz, and definitively retrieved till 3 kHz. This behaviour is not simply related to the duct mode $1a$ solution, which, in fact, shows unstable propagation from 500 till 3 kHz, in case of $M_b=-1.5$. Indeed, both modes $1a$ and $1b$ participate in the scattering problem. In particular, mode $1b$ (check Fig. \ref{fig:cgnANDReImKxVSfreq_mode1_varying_Mb}) restores its passivity around 950 Hz. In order to fully identify the participation of duct modes solutions $1a$ and $1b$ (as well as of higher order modes), a mode-matching analysis will be carried out in a future dedicated study, where all duct-mode solutions will be correlated to a multi-modal scattering problem. In case of $M_b=-1$ instead, mode $1b$ is always a plane wave (stable), therefore it does not affect the passivity of the ABL. Fig.s \ref{fig:ScattCoeff_LEFTexc_VARYINGmuMmuKRat} and \ref{fig:ScattPowCoeff_LEFTexc_hVarying} show the effect of varying the quality factor and the duct cross section width, respectively. The advection speed is fixed with $M_b=-1$, hence, as said before, only the duct mode $1a$ is impacting the passivity in the scattering solution. Indeed, the scattering coefficients of Fig.s \ref{fig:ScattCoeff_LEFTexc_VARYINGmuMmuKRat} and \ref{fig:ScattPowCoeff_LEFTexc_hVarying} perfectly correlate with the modal plots of Fig.s \ref{fig:cgnANDReImKxVSfreq_mode9_varying_QualityFactor} and \ref{fig:cgnANDReImKxVSfreq_mode9_varying_h}, with the loss of passivity confined in a bandwidth starting above $\zeta_{Loc}$ resonance. Fig. \ref{fig:ScattCoeff_LEFTexc_VARYINGmuMmuKRat} confirms that by reducing the quality factor (decreasing $\mu_M=\mu_K$ or augmenting $r_d$) we can restore the acoustical passivity of the ABL in grazing incidence. Therefore, for any duct cross-section width, the scattering solutions confirm the outcomes of modal analysis, according to which it should always be possible to have a passive behaviour of the ABL in the frequency range of interest (here the plane wave regime of the rigid waveguide) by either reducing $M_b$ or the quality factor.\\
Remark that, in the plane wave regime of the hard-walled duct, the scattering solutions give no information about the energy exchanged with higher order rigid-duct modes. Indeed, those latter ones are not able to propagate along the rigid-duct segments preceding and following the liner. Therefore, an apparent passive behaviour of the scattering coefficients in the plane wave regime, is not correlated to an \emph{absolute passivity} as it is defined in Eq. \eqref{eq:absolute acoustical passivity grazing incidence}. Indeed, in order to assess absolute passivity from scattering solutions, we should solve the scattering problem at all frequencies. In the case study reported in this paper, the passive behaviour featured by the ABL in the plane wave regime, is actually related to the \emph{modal} passivity defined in Eq. \eqref{eq:modal acoustical passivity grazing incidence}, relative to modes $1a$ and $1b$. Therefore, in order to assure no amplification of propagated energy in the frequency range of interest, the \emph{modal} passivity criteria defined in Eq. \eqref{eq:modal acoustical passivity grazing incidence} plays an important role. We finally invite the reader to remark that the loss of acoustical passivity always concerns propagation (either forward transmission or backward reflection) in the same direction as $M_b$. This is so, in the open field case of Section \ref{sec:passivity discussion semi-inf domain}, in the duct mode analysis of Section \ref{sec:duct modes analysis}, and in the scattering solution of the present Section.

\begin{itemize}
	\item \textbf{In case of purely real $\zeta_{Loc}$, the scattering coefficients perfectly correlate with the duct mode solutions $1^+$ and $1^-$, in terms of isolation performances, modal passivity and non-reciprocal propagation.}
	\item \textbf{The enhancement of isolation performances induced by the ABL, for excitation field propagating against $M_b$, manifests itself with higher backward reflection, in case of ABL lining both upper and lower sides of the duct.}
	\item \textbf{In case of complex $\zeta_{Loc}$, we have good correlation between the first duct-mode solutions and the scattering coefficients. Nevertheless, because of the change of sign the first duct-mode solutions, a proper mode-matching analysis is needed to perfectly capture the participation of each mode in the scattering performances.}
	\item \textbf{In case of complex $\zeta_{Loc}$, the loss of acoustical passivity related to a reversed direction of duct-mode propagation (change of sign of Re$\{k_x\}$), corresponds to a backward reflection coefficient higher than 1. The unstable propagation always happens in the same sense as $M_b$.}
	\item \textbf{The scattering coefficients confirm the non-reciprocal propagation for $|M_b|=1$.}
\end{itemize}

\FloatBarrier

\section{Scattering simulations in 3D waveguide}\label{sec:scatt 3D}

\begin{figure}[ht!]
		\centering
		\includegraphics[width=\textwidth]{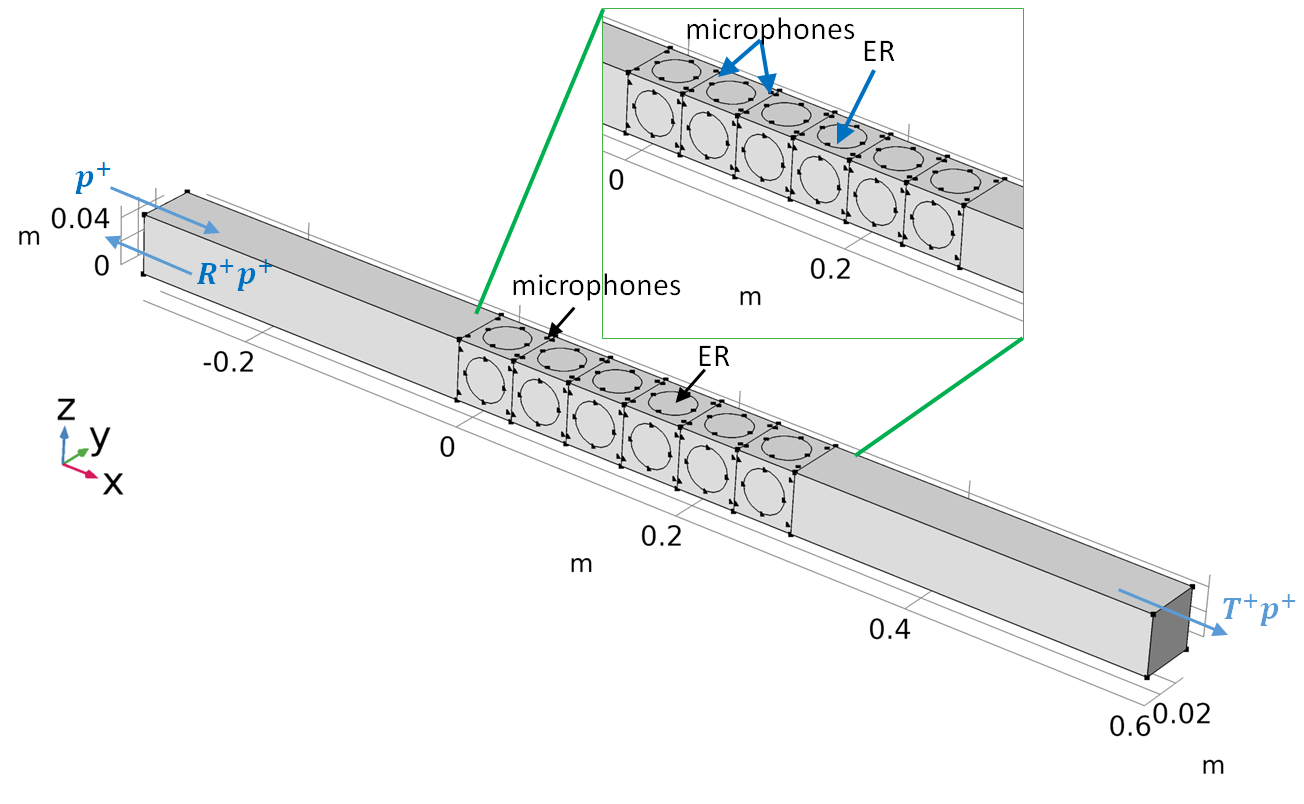}
		\centering
	\caption{3D geometry for scattering simulations, in case of ERs disks applied flush on the duct boundary.}
\label{fig:duct3D_punctulamicros_LSinfront_0.05d_sect}
\end{figure}

\begin{figure}
	\centering
	\includegraphics[width=.6\textwidth]{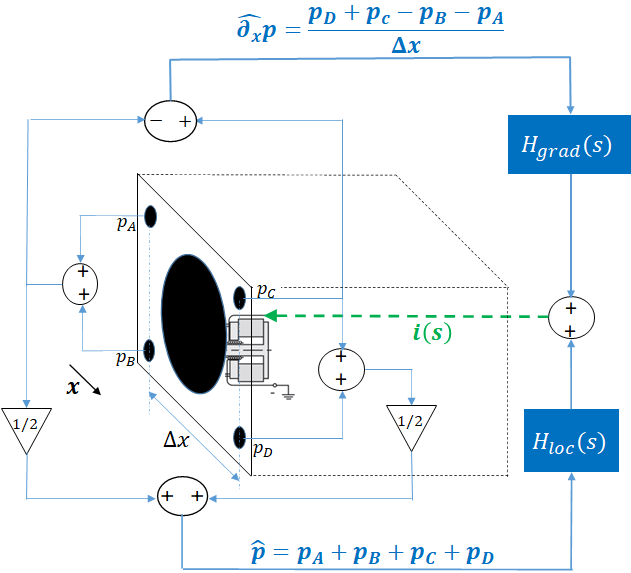}\\
	\caption{Sketch of the 4-microphones ER control, corresponding to Eq. \eqref{eq:controlling current}.}
	\label{fig:loudspeaker_with_NonLocalControl_scheme}
\end{figure}

In this section we simulate the scattering performances in the plane wave regime of a 3D acoustic waveguide, of square cross section with $5$ cm lateral sides, without flow. The ABL is applied along each side of the duct for a length of $30$ cm. In order to investigate the effect of discretizing the ABL by individual ERs lining the parietal walls of a rectangular cross section duct, as in the experimental test-rig of Section \ref{sec:experimental grazing incidence}, the ABL is applied on separate disks of diameter $3.6$ cm, simulating the ERs (6 per each duct edge), as showed in Fig. \ref{fig:duct3D_punctulamicros_LSinfront_0.05d_sect}. The dynamics of each speaker is simulated according to the Thiele-Small SDOF model \cite{beranek2012acoustics}.


The loudspeaker model is reported in Eq. \eqref{eq:loudspeaker model}, in terms of the Laplace variable $s$:

\begin{equation}\label{eq:loudspeaker model}
Z_0(s) \bar{v}(s) = \bar{p}(s) - \frac{Bl}{S_e}\bar{i}(s).
\end{equation}

In Eq. \eqref{eq:loudspeaker model}, $\bar{p}(s)$ and $\bar{v}(s)$ are the acoustic pressure and velocity, respectively, on the speaker diaphragm,  $\bar{i}(s)$ is the electrical current in the speaker coil, $Z_0(s)=M_0s+R_0+K_0/s$ is the acoustical impedance of the loudspeaker in open circuit, with $M_0$, $R_0$ and $K_0$ the corresponding acoustical mass, resistance and stiffness. The electrical current $\bar{i}(s)$ is multiplied by the force factor $Bl$ to get the electromagnetic force, and divided by the effective area $S_e$ to retrieve an equivalent pressure. Observe that the impedance description of Eq. \ref{eq:loudspeaker model} is a lumped-element model, which is reliable as long as the wavelength of the acoustic field is sufficiently larger than the size of the speaker diaphragm. This is true for any local impedance modelling. The upper frequency of validity of the lumped-element assumption is much beyond the frequency range of validity of the SDOF loudspeaker-model, which lies around the first speaker mode (around 468 Hz). Therefore, both the lumped-element assumption and the SDOF model are valid around the principal resonance of the ER.\\
The ABL is implemented by defining the electrical current $i(s)$ as in Eq. \eqref{eq:controlling current}:

\begin{equation}
	\label{eq:controlling current}
	i(s)  = H_{loc}(s)\hat{\bar{p}}(s) + H_{grad}(s)\hat{\partial_x} \bar{p}(s),
\end{equation}

where $\hat{\bar{p}}(s)$ and $\hat{\partial_x} \bar{p}(s)$ are the estimated local pressure and its x-derivative on each speaker diaphragm, in the Laplace domain. The local sound pressure is estimated by averaging the four microphones on the corners of each ER $\hat{p}=(p_A+p_B+p_C+p_D)/4$, while the x-derivative is estimated by a first-order finite difference $\hat{\partial_x} p=\biggr((p_C+p_D)-(p_A+p_B)\biggr)/\Delta x$, with $\Delta x\approx 4$ cm the distance between the microphones before (A,B) and after (C,D) each ER speaker, along the x-direction, as showed in Fig. \ref{fig:loudspeaker_with_NonLocalControl_scheme}. A time delay of $\tau=2\times10^{-5}$ seconds between the pressure inputs and the electrical current, is considered by multiplying the microphones pressures by $e^{-\mathrm{j}\omega\tau}$, in order to simulate the physiological latency of the digital control algorithm of the ER \cite{de2022effect}.\\
The transfer functions $H_{loc}(s)$ and $H_{grad}(s)$ are defined based upon the loudspeaker model of Eq. \eqref{eq:loudspeaker model}. Equating the velocity of the speaker diaphragm from Eq. \eqref{eq:loudspeaker model}, and the velocity corresponding to the ABL (Eq. \eqref{eq:advection law in vy with Z}), we get the expressions in the Laplace space of $H_{loc}$ and $H_{grad}$, in Eq.s \eqref{eq:Hloc} and \eqref{eq:Hgrad}, respectively.

\begin{equation}
	\label{eq:Hloc}
	H_{loc}(s) = \frac{S_e}{Bl} \biggr(1 - \frac{Z_{0}(s)}{Z_{Loc}(s)}\biggr),\\
\end{equation}

\begin{equation}
	\label{eq:Hgrad}
	H_{grad}(s) = -\frac{S_e}{Bl}\frac{Z_{0}(s)}{Z_{Loc}(s)}\frac{U_b}{s}F_{hp}(s),\\
\end{equation}

where $F_{hp}(s)$ in $H_{grad}(s)$ is a high-pass filter necessary in order for $H_{grad}(\mathrm{j}\omega)$ not to become infinite for $\omega\to 0$. Notice also, that a purely real $Z_{Loc}$ would lead to non-causal $H_{loc}$ and $H_{grad}$, therefore we have employed the SDOF expression of Eq. \eqref{eq:zeta_Loc complex} for $Z_{Loc}$ (as in Sections \ref{sec:duct modes complex impedance} and \ref{sec:scatt complex impedance}) in the correctors $H_{loc}$ and $H_{grad}$.
The synthesis of our corrector transfer functions, is also called \emph{model inversion} \cite{devasia2002should} approach, as the objective of the controller is to cancel out the loudspeaker proper dynamics, and replace it with a desired acoustic behaviour. Both $H_{loc}$ and $H_{grad}$ depends upon the loudspeaker own impedance model $Z_0$. Therefore, each parameter appearing in Eq. \eqref{eq:loudspeaker model} must be estimated. The so-called Thiele-Small parameters are identified by acoustic measurements, as described in \cite{debono2019electroacoustic}, and their values are reported in Table \ref{tab:TSparam FEMTO}. Further details upon such control strategy can be found in \cite{de2022effect,DeBono2024}.\\
Both Eq.s \eqref{eq:Hloc} and \eqref{eq:Hgrad} are implemented in the Comsol model. From the microphones estimation of $\hat{p}$ and $\hat{\partial_x} p$, the electrical current $i$ is obtained from Eq. \eqref{eq:controlling current}. Hence, the loudspeaker dynamics Eq. \eqref{eq:loudspeaker model} is solved for $\bar{v}(s)$, which is then imposed on the disks representing the speaker membranes in the numerical model.\\
It is worthy to note that the control filters presented here in Eq.s \eqref{eq:Hloc} and \eqref{eq:Hgrad} target a resonant $Z_{Loc}$, while in references \cite{collet2009active} and \cite{KarkarDeBono2019}, $Z_{Loc}$ was considered as just a mass term, limiting its applicability to frequencies above the loudspeaker resonance.\\
The model showed in Fig. \ref{fig:duct3D_punctulamicros_LSinfront_0.05d_sect} is solved for the scattering coefficients as in Section \ref{sec:scattering 2D}. The FE mesh elements have the same maximum size as those in Section \ref{sec:scattering 2D}.\\

\begin{figure}[ht!]
	\centering
	\begin{subfigure}[ht!]{0.45\textwidth}
		\centering
		\includegraphics[width=\textwidth]{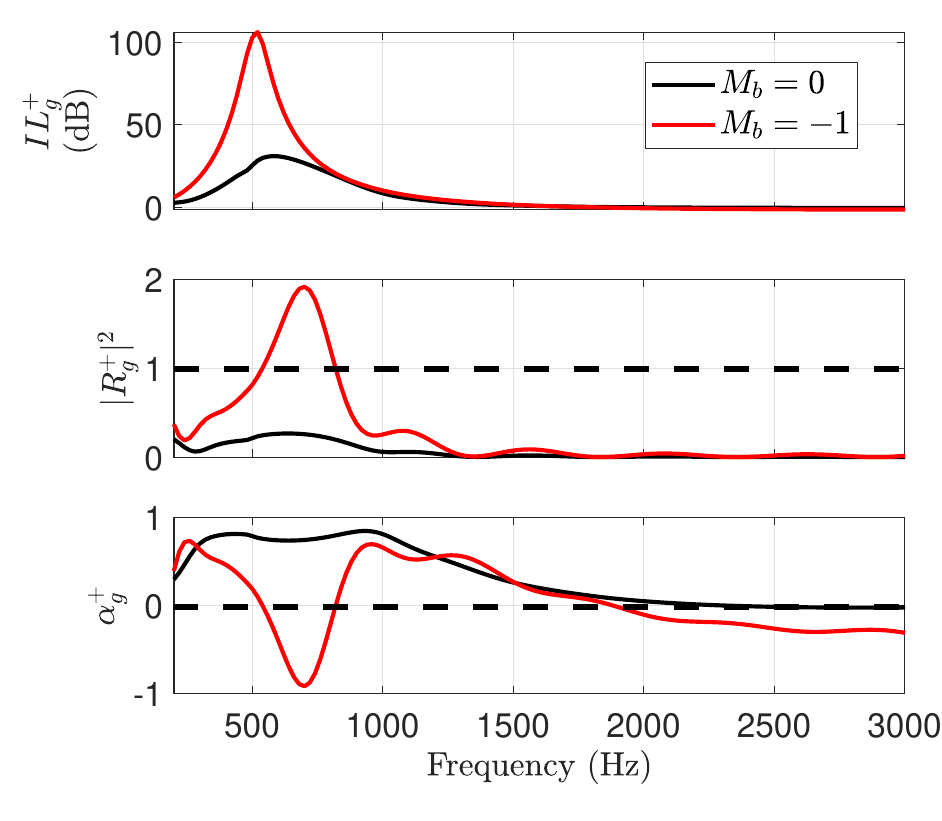}
		\centering
		\caption{}
		\label{fig:Scatt3D_NominalCase_0.5muKmuM_1rho0c0Rat_-1cac0_2e-05tau_OCvsLocalvsNonLocal}
	\end{subfigure}
	\hspace{1 cm}
	\begin{subfigure}[ht!]{0.45\textwidth}
		\centering
		\includegraphics[width=\textwidth]{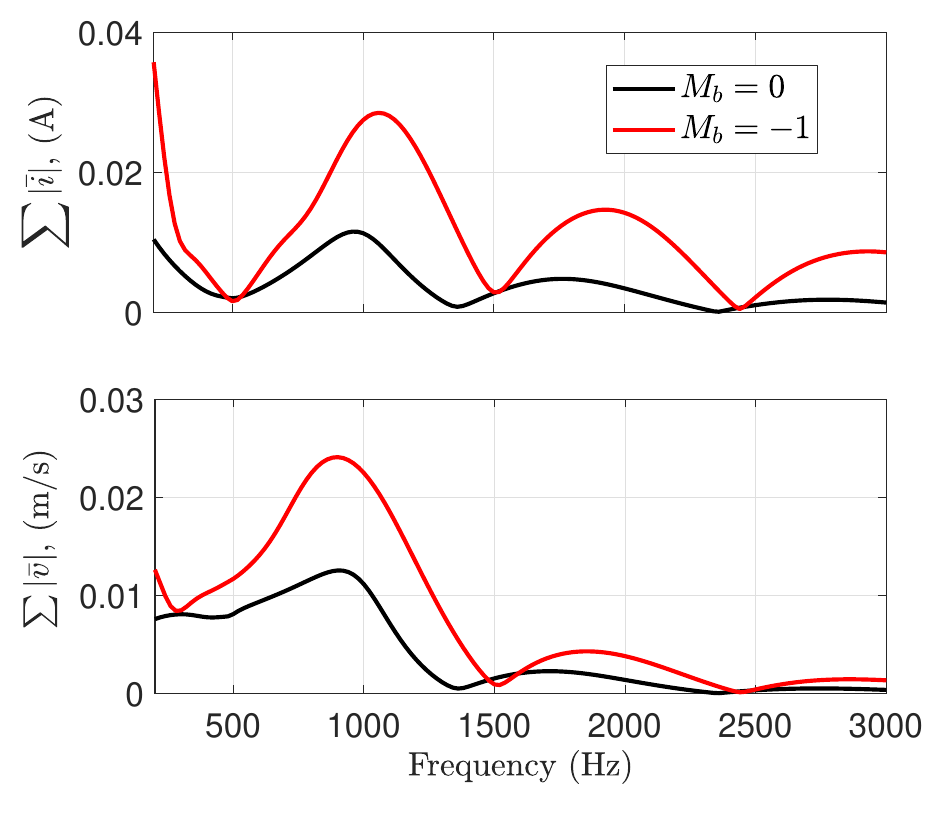}
		\centering
		\caption{}
		\label{fig:curr_vel_TotalAllCells_3D_NonLocalVSLocal_0.5muKmuM_1rho0c0Rat_2e-05tau}
	\end{subfigure}
	\caption{Comparison between local impedance control ($M_b=0$) and ABL ($M_b=-1$), in terms of scattering coefficients in the 3D waveguide \textbf{(a)}, and in terms of sum of all ERs electrical current spectra $\sum |\bar{i}|$ and velocity spectra $\sum |\bar{v}|$ \textbf{(b)}.}
	\label{fig:Scatt3D NominalCase}
\end{figure}

In Fig. \ref{fig:Scatt3D_NominalCase_0.5muKmuM_1rho0c0Rat_-1cac0_2e-05tau_OCvsLocalvsNonLocal}, the scattering coefficients achieved by the ABL with $M_b=-1$, are plotted along with the ones relative to local impedance control ($M_b=0$), applied on each ER. The $\zeta_{Loc}$ parameters are set to $\mu_M=\mu_K=0.5$ and $R_d=\rho_0c_0$. As in the 2D case, the ABL demonstrates higher isolation capabilities, though being non-passive slightly after resonance. Notice also the loss of passivity at high frequencies (above 2 kHz), which was not predicted by the 2D simulations. This is mostly due to the time delay \cite{de2022effect}. In Appendix \ref{app:discretization and delay effect}, we briefly check the effects of the finite difference approximation of $\hat{\partial_x} p$ and of time delay. In Fig. \ref{fig:curr_vel_TotalAllCells_3D_NonLocalVSLocal_0.5muKmuM_1rho0c0Rat_2e-05tau}, the electrical current spectra of all the ERs are summed up to visualize how the ABL requires a much higher level of electrical current (up to 3 times at some frequencies) with respect to the local impedance control. Also the sum of velocities on the 24 ERs is reported showing once again higher vibrational amplitudes required by the ABL.\\

\begin{figure}[ht!]
	\centering
	\begin{subfigure}[ht!]{0.45\textwidth}
		\centering
		\includegraphics[width=\textwidth]{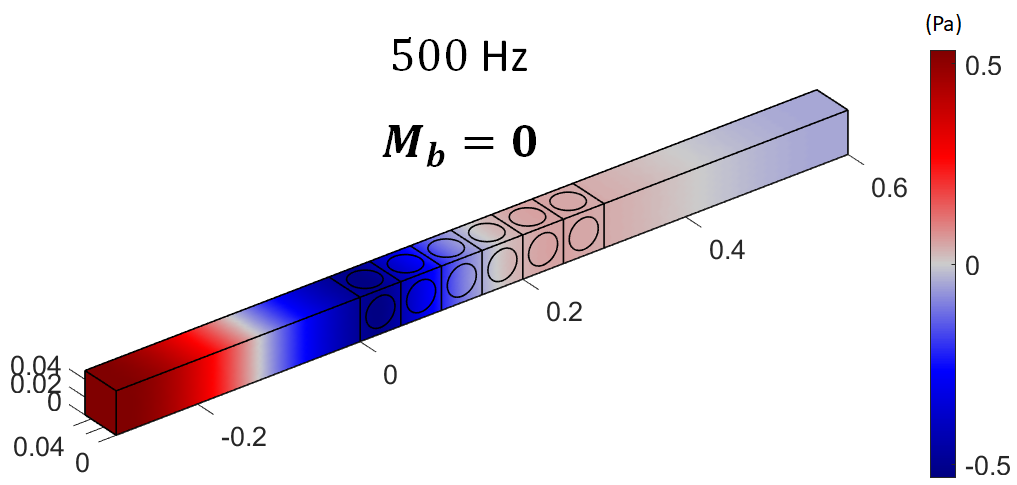}
		\centering
		\caption{}
		\label{fig:surfaceplot3D_LocalControl_500Hz}
	\end{subfigure}
	\hspace{0.5 cm}
	\begin{subfigure}[ht!]{0.45\textwidth}
		\centering
		\includegraphics[width=\textwidth]{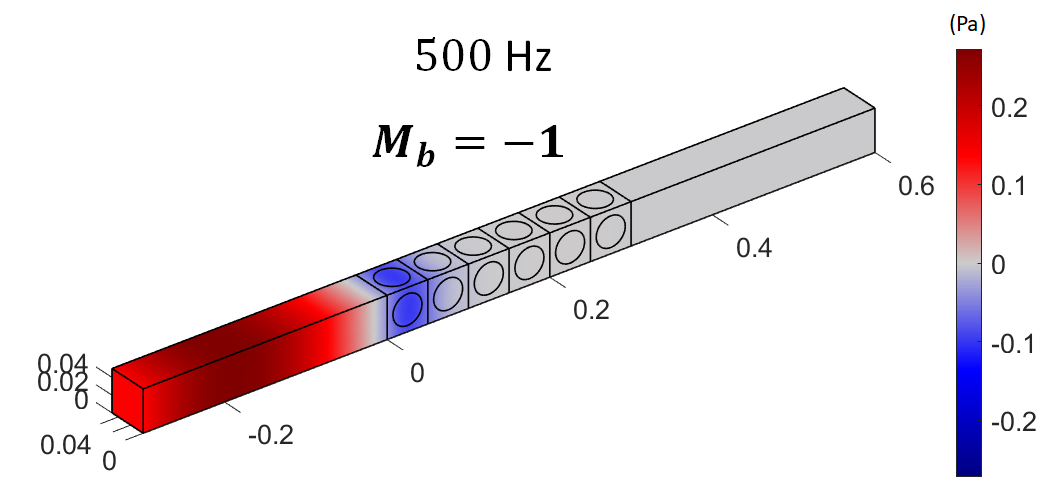}
		\centering
		\caption{}
		\label{fig:surfaceplot3D_NonLocalControl_500Hz}
	\end{subfigure}
	\caption{3D surface plots of the sound pressure field at 500 Hz, in case of local impedance control \textbf{(a)} or ABL with $M_b=-1$ \textbf{(b)}, on the ERs.}
	\label{fig:3D surf plots}
\end{figure}

Fig. \ref{fig:3D surf plots} compares the sound field at 500 Hz computed in the duct for an incident pressure of 1 Pa, when local impedance control (Fig. \ref{fig:surfaceplot3D_LocalControl_500Hz}) or ABL (Fig. \ref{fig:surfaceplot3D_NonLocalControl_500Hz}) is applied on the ERs. Observe how the sound pressure field gets annihilated as soon as it enters the segment lined by the ABL. The enhancement of sound transmission attenuation for $M_b=-1$, with respect to the case of $M_b=0$, is unequivocal.\\

\FloatBarrier

\section{Experimental results}\label{sec:experimental grazing incidence}

\begin{figure}[ht!]
	\centering
	\includegraphics[width=\textwidth]{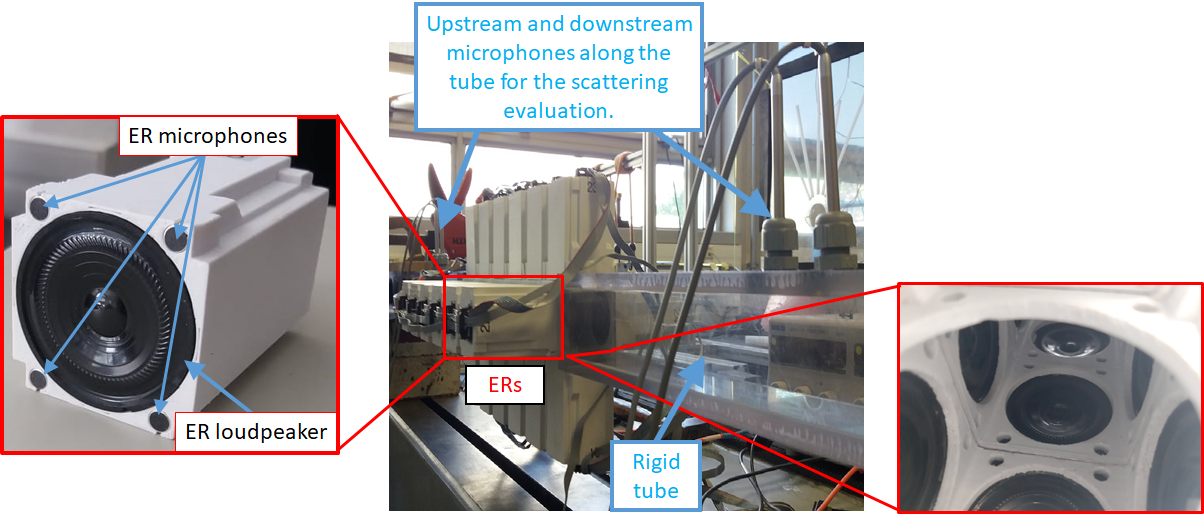}
	\caption{ER prototype (left); waveguide (middle) for the scattering evaluation, with internal view of the lined segment (right).}
	\label{fig:TL_ExpSetup}
\end{figure}

\begin{figure}[ht!]
	\centering
	\includegraphics[width=\textwidth]{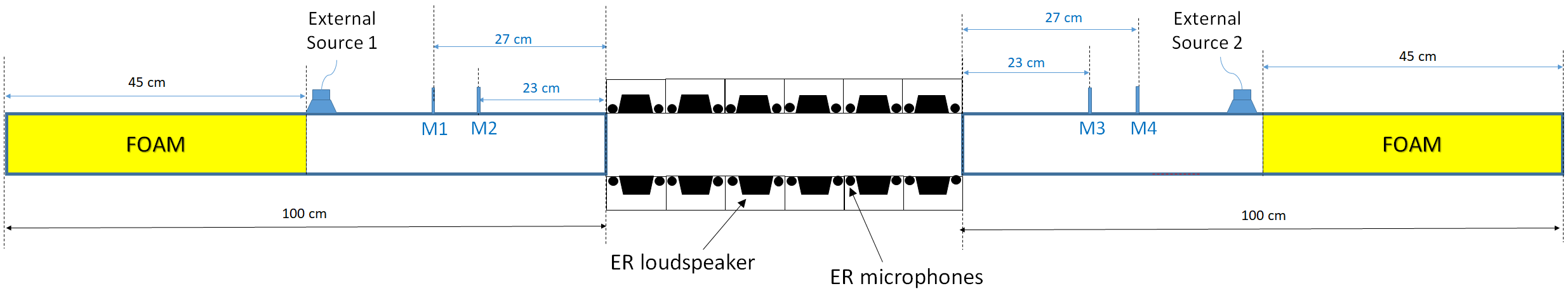}
	\caption{Sketch of the test-bench.}
	\label{fig:TL_Setup_Sketch}
\end{figure}

\begin{figure}
	\centering
	\includegraphics[width=.6\textwidth]{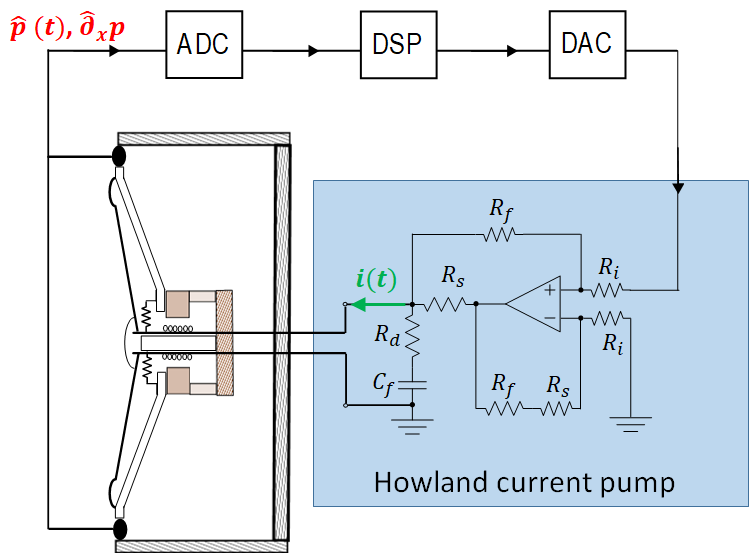}
	\caption{Sketch of the ER architecture.}
	\label{fig:EA_with_Howland}
\end{figure}

In this section, the advection control law is experimentally tested on an array of 24 ER prototypes lining a squared cross-section duct of about $0.05$ m side, as illustrated in the photos of Fig. \ref{fig:TL_ExpSetup} and in the sketch of Fig. \ref{fig:TL_Setup_Sketch}. The ERs are placed 6 per each side of the duct, as showed in Fig. \ref{fig:TL_ExpSetup}. Each ER has a surface area of about $0.05\times0.05$ $m^2$, for a total lined segment length of about $0.3$ m in the duct. Both ends of the tube are filled with 45 cm of foam to reproduce quasi-anechoic conditions at the input and output of the waveguide. An external acoustic source is placed flush with the duct surface on both sides of the waveguide, just ahead of the foam terminations, sufficiently far from the lined segment and from microphone locations. The external sources are excited with a sine-sweep signal from 150 Hz (lower limit of the source-loudspeakers) to 3 kHz (to stay below the cut-on frequency of the higher rigid duct modes).\\
Each ER is controlled autonomously, and the control architecture is illustrated in Fig. \ref{fig:EA_with_Howland}: the signals $\hat{p}$ and $\hat{\partial}_xp$ on the speaker diaphragm, after being digitally converted by the Analogue-Digital-Converter (ADC), are fed into a \emph{programmable} digital signal processor (DSP) where the output of the control is computed at each time step. The Howland current pump \citep{pease2008comprehensive} allows to enforce the electrical current $i$ in the speaker coil independently of the voltage at the loudspeaker terminals. It consists of an operational amplifier, two input resistors $R_i$, two feedback resistors $R_f$, and a current sense resistor $R_s$. The resistance $R_d$ and capacitance $C_f$ constitutes the compensation circuit to ensure stability with the grounded load \citep{steele1992tame}. More details can be found in \cite{de2022effect}.

All ERs and control interfaces have been produced in the Department of Applied Mechanics at FEMTO-st Institute. The control laws have already been defined in Section \ref{sec:scatt 3D}, by Eq. \eqref{eq:controlling current}, \eqref{eq:Hloc}, \eqref{eq:Hgrad}, and the loudspeaker parameters provided in Table \ref{tab:TSparam FEMTO}. The four scattering coefficients have been estimated according to the two-source method \cite{munjal1990theory}.

\begin{figure}[ht!]
	\centering
	\includegraphics[width=0.6\textwidth]{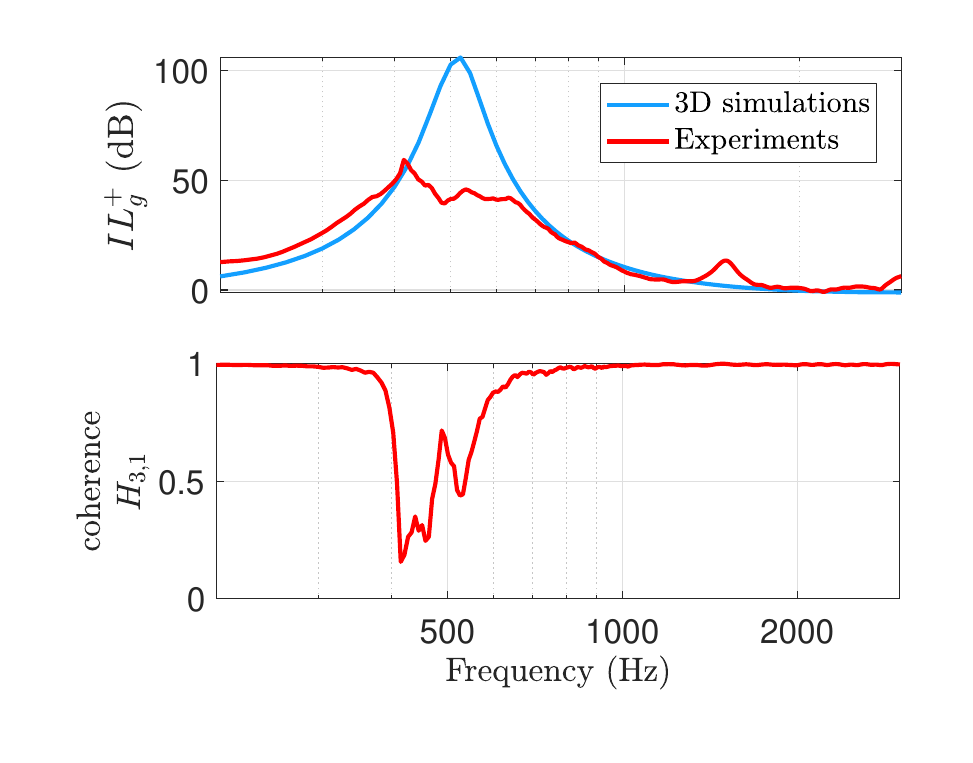}
	\centering
	\caption{Comparison between measurements (in red) and simulations (in blue) for ABL with $\mu_M=\mu_K=0.5$, $R_{d}=\rho_0c_0$ and $M_b=-1$.}
	\label{fig:measVSsimul}
\end{figure}

In Fig. \ref{fig:measVSsimul}, the $IL_g^+$ from measurement is compared to the one obtained from 3D simulations (given in Section \ref{sec:scatt 3D}), for $\mu_M=\mu_K=0.5$, $R_{d}=\rho_0c_0$ and $M_b=-1$. Observe how, despite the inevitable model uncertainties, the trends before and after the resonance peak are well captured by the 3D simulations, except around 1.5 kHz where an additional speaker mode appears, as in \cite{de2022effect}. The peak of more than 100 dB of attenuation predicted by the simulations, is not visible experimentally. This is indeed due to the very low signal-to-noise ratio at microphone 3 caused by the extreme isolation accomplished by the ABL. This prevents the detection of very high $IL$ values, as confirmed by the low level of coherence around resonance, of the transfer functions between microphones on opposite sides with respect to the lined segment (check the coherence of transfer function $H_{3,1}$ between microphones 3 and 1, in Fig. \ref{fig:measVSsimul}).

\begin{figure}
	\centering
	\begin{subfigure}[ht!]{0.45\textwidth}
		\centering
		\includegraphics[width=\textwidth]{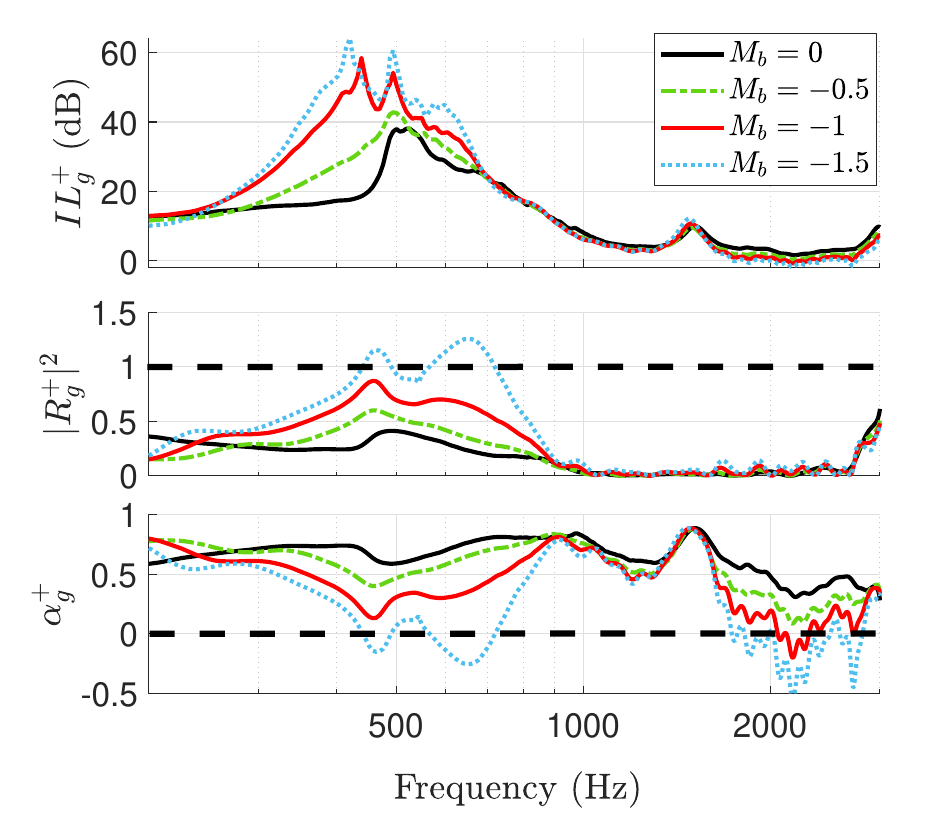}\\
		\caption{}
		\label{fig:ExpSCATTplus_VARYINGcac0_0.5muM_1rho0c0Rat_0.5muK}
	\end{subfigure}	
\hspace{0.5 cm}
	\begin{subfigure}[ht!]{0.45\textwidth}
		\centering
		\includegraphics[width=\textwidth]{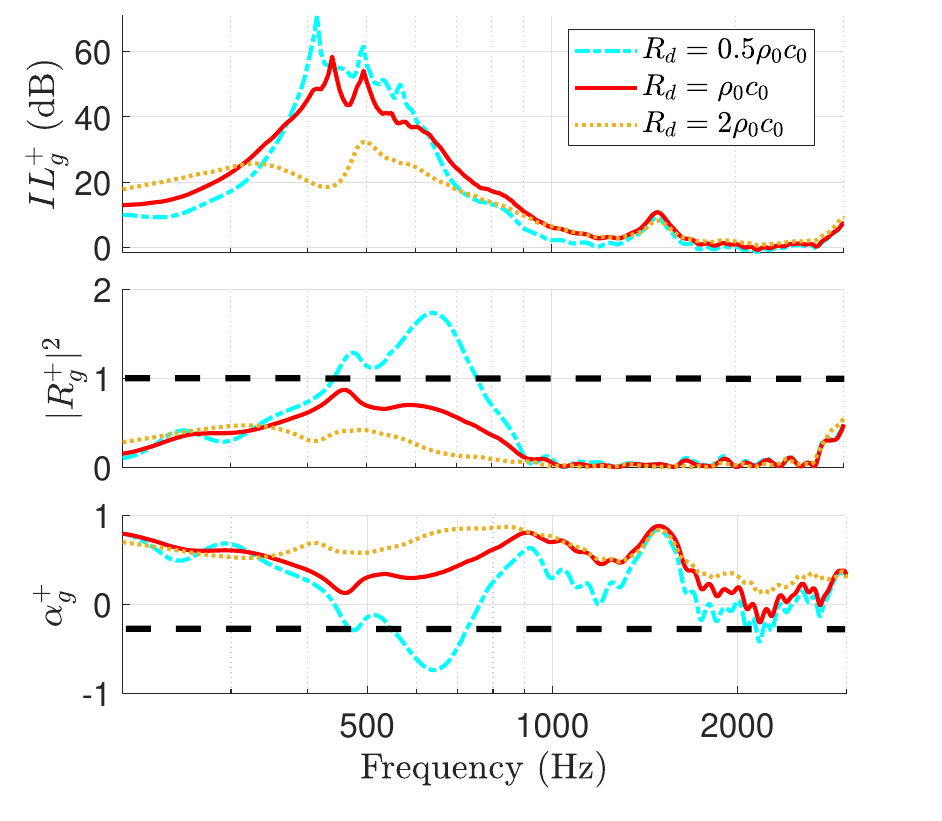}\\
		\caption{}
		\label{fig:ExpSCATTplus_-1cac0_0.5muM_VARYINGrho0c0Rat_0.5muK}
	\end{subfigure}
	\caption{Experimental scattering performances for incident field propagating toward $+x$, achieved by the ABL with varying $M_b$ (\textbf{a}), or varying $R_d$. The default parameters are set to $\mu_M=\mu_K=0.5$, $R_{d}=\rho_0c_0$ and $M_b=-1$.}
	\label{fig:ExpScatt_varying_MbRat}
\end{figure}

Fig. \ref{fig:ExpSCATTplus_VARYINGcac0_0.5muM_1rho0c0Rat_0.5muK} and \ref{fig:ExpSCATTplus_-1cac0_0.5muM_VARYINGrho0c0Rat_0.5muK} show the experimental scattering coefficients for incident field toward $+x$, with varying $M_b$ and $R_d$ respectively. The default parameters are set to $\mu_M=\mu_K=0.5$, $R_{d}=\rho_0c_0$ and $M_b=-1$. Fig. \ref{fig:ExpSCATTplus_VARYINGcac0_0.5muM_1rho0c0Rat_0.5muK} confirms the higher isolation achieved by increasing the absolute value of $M_b<0$, though the $IL_g^+$ for $M_b=-1.5$ does not look significantly augmented with respect to $M_b=-1$. This, once again, can be explained by an excessively low signal-to-noise ratio of microphones after the lined segment (microphones 3 and 4 for positive propagation), and the consequent low coherence of the corresponding transfer function. The reflection and absorption coefficients though, are still able to follow the expected trends, with the loss of passivity immediately after resonance. Fig. \ref{fig:ExpSCATTplus_-1cac0_0.5muM_VARYINGrho0c0Rat_0.5muK} also validates the numerical predictions both in terms of isolation performances and passivity, demonstrating that increasing the quality factor brings about an excess in the backward reflection, endangering passivity above resonance. Observe, in Fig. \ref{fig:ExpSCATTplus_VARYINGcac0_0.5muM_1rho0c0Rat_0.5muK}, the reduction of passivity from 1.8 kHz and above with higher $|M_b|$. This is due to a combined effect of time delay and the first order approximation of $\hat{\partial}_x p$, which is clearly amplified for higher values of $|M_b|$.

\begin{figure}[ht!]
	\centering
	\includegraphics[width=0.6\textwidth]{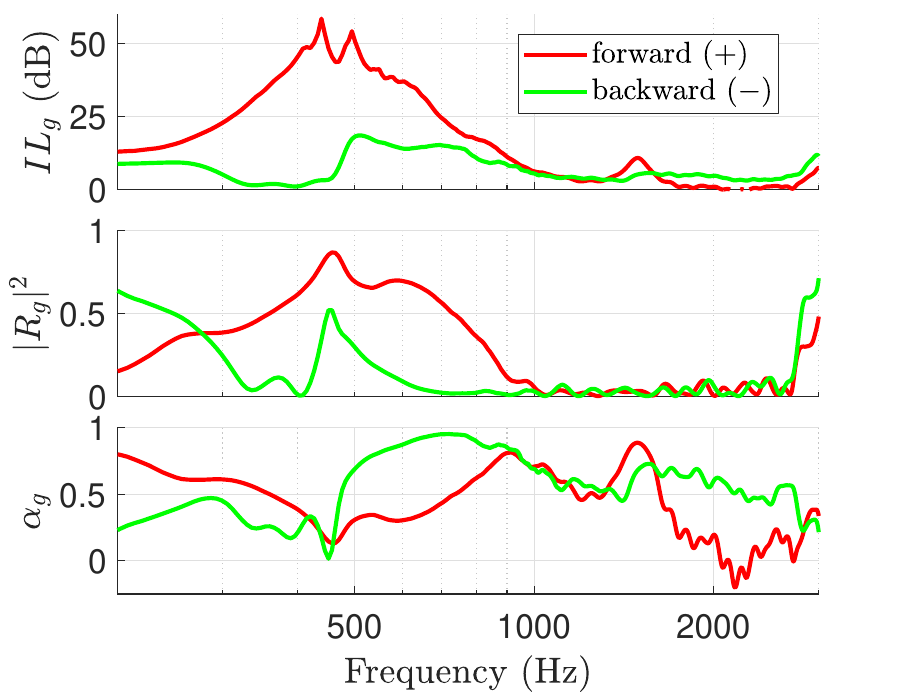}
	\caption{Scattering performances relative to external incident field propagating toward $+x$ (``forward'', in solid red) compared to the ones relative to ``backward'' incident field (in solid green), in case of ABL with $\mu_M=\mu_K=0.5$, $R_{at}=\rho_0c_0$ and $M_b=-1$.}
	\label{fig:ExpSCATTplusANDminus_NonReciprocity}
\end{figure}

The broadband non-reciprocal character of the advective BC is evident by looking at Fig. \ref{fig:ExpSCATTplusANDminus_NonReciprocity}, where the ``forward'' scattering coefficients (corresponding to the first column of the scattering matrix of Eq. \eqref{eq:scattering matrix definition}), are plotted along with the ``backward'' scattering coefficients (corresponding to the second column of the scattering matrix of Eq. \eqref{eq:scattering matrix definition}), in case of $\mu_M=\mu_K=0.5$, $r_d=1$ and $M_b=-1$.
Observe that, in the backward direction, we do not have perfect transmission. Indeed, because of time-delay and model-uncertainties in the actual control system, the model-based correctors $H_{loc}$ and $H_{grad}$ of Eq. \eqref{eq:Hloc} and \eqref{eq:Hgrad} are not capable to fully cancel out the actuator dynamics, leading to residual non-zero loudspeaker response and non-perfect transmission in the backward direction. Further details on the limitation of such corrector synthesis approach can be found in \cite{de2022effect}. Nevertheless, $IL_g^-$ never overcomes 18 dB, while for forward propagation $IL_g^+$ is significantly higher than 25 dB from 300 to 700 Hz, and higher than 50 dB close to resonance. Notice that such non-reciprocal propagation is achieved in the bandwidth of $\zeta_{Loc}$, while in \cite{KarkarDeBono2019} it was accomplished only above resonance. This is due to the different definitions of $H_{loc}$ and $H_{grad}$, which are here targeting the frequency range around $\zeta_{Loc}$ resonance, allowing to significantly enhance both isolation and non-reciprocal performances in the target bandwidth.\\

Below, the main outcomes of Sections \ref{sec:scatt 3D} and \ref{sec:experimental grazing incidence}.

\begin{itemize}
	\item \textbf{The corrector transfer functions are synthesized based upon the model-inversion strategy.}
	\item \textbf{The ABL is enforced by estimating the average and the first $x$-derivative of $p$, on each ER.}
	\item \textbf{The 3D scattering simulations confirm higher isolation capabilities of the ABL with respect to local impedance operators.}
	\item \textbf{The experimental implementation also confirms the expected results, in terms of isolation, passivity and non-reciprocal propagation.}
	\item \textbf{The physiological time-delay in the digital implementation of the control, endangers high-frequency passivity.}
\end{itemize}

\FloatBarrier

\section{Conclusions}\label{sec:conclusions}

In this article we have provided a detailed discussion of the Advection Boundary Law, which is composed of a local impedance component and a convective term. Starting from its theoretical conception of Section \ref{sec:theoretical conception}, such operator is defined, for the first time, as a degeneration of a Dirichelet-to-Neumann mapping of a semi-infinite non-isotropic propagative domain, on the boundary. As the surface impedance operator can be seen as a special case of Advection Boundary Law, the general framework originally
employed by Morse \cite{morse1939some} to introduce the surface impedance concept, is here retrieved and generalized to include our advective boundary operator. The semi-infinite approach naturally leads to the open-field scattering problem and the corresponding reflection coefficient formula (Section \ref{sec:passivity discussion semi-inf domain}). 
The open-field solution allows the definition of acoustical passivity in open field which, in case of our Advection Boundary Operator, depends upon the incidence angle. 
Following a step-by-step increase in complexity, we discuss the duct-mode solutions in a 2D waveguide without flow in Section \ref{sec:duct modes analysis}. In particular, we first analyse the case of purely real local impedance term (Section \ref{sec:duct modes real impedance}), and then the case of complex local impedance term (Section \ref{sec:duct modes complex impedance}). The duct mode analysis leads to the distinction between \emph{absolute} and \emph{modal} passivity. In case of purely real local impedance, the passivity limits in open-field assure the absolute passivity in grazing incidence. Nevertheless, in case of reactive local impedance, the passivity limits in open-field do not assure absolute passivity in grazing incidence. In particular, for a boundary advection speed against the incident field, modal passivity is affected by both the reactive component of the local impedance, and the boundary advection speed. Moreover, such impact is stronger for narrower ducts. Nevertheless, for any duct-cross section sizes, it is always possible to restore stability of the duct-modes of interest, i.e. to assure the corresponding \emph{modal} acoustical passivity of our Advection Boundary Law. The 2D duct-mode analysis is followed by the resolution of the 2D scattering problem. The correlation between the two studies is evident, in terms of passivity, attenuation levels and non-reciprocal propagation. A future study will be dedicated to the resolution of the scattering problem by mode-matching techniques, allowing to visualize numerically the correlation between modal and scattering solutions.\\
In order to guarantee no amplification of propagated energy in the frequency range of interest, the modal passivity plays an important role. In this paper, we have provided a physical quantity able to assess both acoustical passivity limits and attenuation levels, in Section \ref{sec:duct modes analysis}. It is the sine of the elevation angle of modal local group velocity at the boundary. Such quantity can be employed for liner optimization purposes. Moreover, its nice physical interpretation, allows to clarify the mechanism leading to enhanced attenuation achieved by the Advection Boundary Law, and should be taken into account for the design of next generation liners.\\
The final step of complexity in the numerical simulations, is the 3D scattering solution provided in Section \ref{sec:scatt 3D}, where the Advection Boundary Law is discretized and implemented on Electroacoustic Resonators, composed of a loudspeaker and four microphones. The 3D scattering results confirm that the enhanced isolation performances are still achieved despite the boundary discretization, and provide an intermediate step before the experimental validation of Section \ref{sec:experimental grazing incidence}. An array of programmable Electroacoustic Resonators lining an acoustic waveguide allows to implement the Advection Boundary Law in real life. The measurements validate the Advection Boundary Law accomplishments in terms of enhanced isolation, passivity and non-reciprocal sound propagation, despite the physiological limitations of digital control algorithms.\\
Because of its non-natural and non-local character, special attention must be given when implementing the Advection Boundary Law. In this paper, we have provided a range of interpretational and numerical tools to guide the control users when implementing such special boundary control, in order to maximize its isolation performances, avoid non-passive behaviours, and/or achieve the desired non-reciprocal propagation. This first study has analysed the Advection Boundary Law in the plane-wave regime and in absence of mean flow. Such work has put the necessary bases for the Advection Boundary Law to tackle more complex guided propagation problems, including airflow convection and multi-modal propagation. 

\FloatBarrier

\begin{appendices}

\section{Duct modes problem formulation}\label{app:Duct modes problem formulation}
Consider an infinite duct of constant cross-section $\mathcal{A}$ in the plane $y,z$ (as in Fig. \ref{fig:arbitrary_duct}) with boundary $\partial\mathcal{A}$ and normal $\vec{n}$. Assuming a time-harmonic sound field in the usual complex notation ($+\mathrm{j}\omega t$) in the duct, the wave equation reduces to the Helmholtz equation:

\begin{equation}
	\label{eq:Helmholtz equation}
	\nabla^2\bar{p} + k_0^2\bar{p} = 0.
\end{equation}

Such sound field must also satisfy the generic BC $\mathcal{B}(\bar{p})=0$ on the wall $\partial\mathcal{A}$. The solution to this problem can be written as:

\begin{equation}
	\label{eq:solution sum of duct modes}
	\bar{p}(t,\omega,x,y,z) = e^{\mathrm{j}\omega t}\sum_{m=0}^{\infty} A_m \psi_{m}(\omega,y,z)e^{-\mathrm{j}k_{x,m}(\omega)x},
\end{equation}

where $\psi_{m}(y,z)$, the so-called \emph{duct modes}, are the eigenfunctions of the transverse Laplace operator reduced to $\mathcal{A}$ satisfying the BC $\mathcal{B}[p]=0$ on $\partial\mathcal{A}$, i.e. they are solution of the eigenvalue problem:

\begin{equation}
	\label{eq:duct modes problem}
	\begin{split}
		&\nabla^2_{yz} \psi_m(y,z) + (k_0^2 - k_{x,m}^2)\psi_m(y,z)=0 \;\;\; \mathrm{for \; y,z \in }\; \mathcal{A}\\
		&\mathcal{B}[\psi_m(y,z),k_{x,m}]=0 \;\;\; \mathrm{for \; y,z \in }\; \partial\mathcal{A},
	\end{split}
\end{equation}

where $\nabla^2_{yz}$ denotes the Laplacian operator in $y,z$ (following the notation of \cite{rienstra2015fundamentals}), whose eigenvectors and eigenvalues are the duct mode shapes $\psi_{m}(y,z)$ and ($k_0^2 - k_{x,m}^2$), respectively. Observe that for classical liners, the BC does not involve the axial wavenumber $k_{x,m}$.

We now formulate the duct mode problem in case of ABL as BC, in which the locally reacting liner is a special case (for $M_b=0$). The duct-modes eigenvalue problem writes:

\begin{subequations}
	\label{eq:duct modes problem with advection BC}
\begin{equation}
		\nabla^2_{y,z} \psi_m(y,z) - (k_0^2 - k_{x,m}^2)\psi_m(y,z)=0 \quad \mathrm{for} \;  y \in \mathcal{A}\\
\end{equation}
\begin{equation}
		\vec{n}\cdot\vec{\nabla}\psi_{m}(y,z) = - \mathrm{j}\eta_{Loc}\biggr(k_0 - M_b k_{x,m}\biggr) \psi_{m}(y,z) \quad \mathrm{for} \;  y \in \partial\mathcal{A}.
\end{equation}
\end{subequations}

Notice the \emph{non-standard} character of such eigenvalue problem, where the eigenvalue appears in the BC as well. Solutions for such eigenvalue problem can be found by FEs. The weak formulation of the eigenvalue problem of Eq.s \eqref{eq:duct modes problem with advection BC} is reported in Eq.s \eqref{eq:weak formulation}, where $\hat{\psi}$ is the test function for the duct mode $\psi_m$. The integration by parts (application of Green formula) is given in Eq. \eqref{eq:weak formulation integration by parts}, and the final expression, with the assimilation of our BC, in Eq. \eqref{eq:weak formulation final expression}. Hence, our eigenvalue problem with an eigenvalue-dependent BC, can be solved directly in its weak-form, by FEs.\\

\begin{subequations}
	\label{eq:weak formulation}
	\begin{equation}
		\int_{\mathcal{A}}\hat{\psi}\nabla^2_{y,z} \psi_m \; dydz + (k_0^2 - k_{x,m}^2)\int_{\mathcal{A}}\hat{\psi}\psi_m \; dydz=0
	\end{equation}
	\begin{equation}\label{eq:weak formulation integration by parts}
		\int_{\partial\mathcal{A}} \hat{\psi}\partial_n\psi_m\;dydz-\int_{\mathcal{A}}\vec{\nabla}_{y,z}\hat{\psi}\cdot\vec{\nabla}_{y,z} \psi_m \;dydz + (k_0^2 - k_{x,m}^2)\int_{\mathcal{A}}\hat{\psi}\psi_m\;dydz=0
	\end{equation}
	\begin{equation}\label{eq:weak formulation final expression}
		- \mathrm{j}\eta_{Loc}\biggr(k_0 - M_bk_{x,m}\biggr)\int_{\partial\mathcal{A}} \hat{\psi}\psi_m\;dydz-\int_{\mathcal{A}}\vec{\nabla}_{y,z} \hat{\psi} \cdot\vec{\nabla}_{y,z}\psi_m\;dydz + (k_0^2 - k_{x,m}^2)\int_{\mathcal{A}}\hat{\psi}\psi_m\;dydz=0
	\end{equation}
\end{subequations}

\FloatBarrier

\FloatBarrier

\section{Effect of discrete pressure evaluation and time delay in the 3D numerical model}\label{app:discretization and delay effect}

\begin{figure}[ht!]
	\centering
	\includegraphics[width=.5\textwidth]{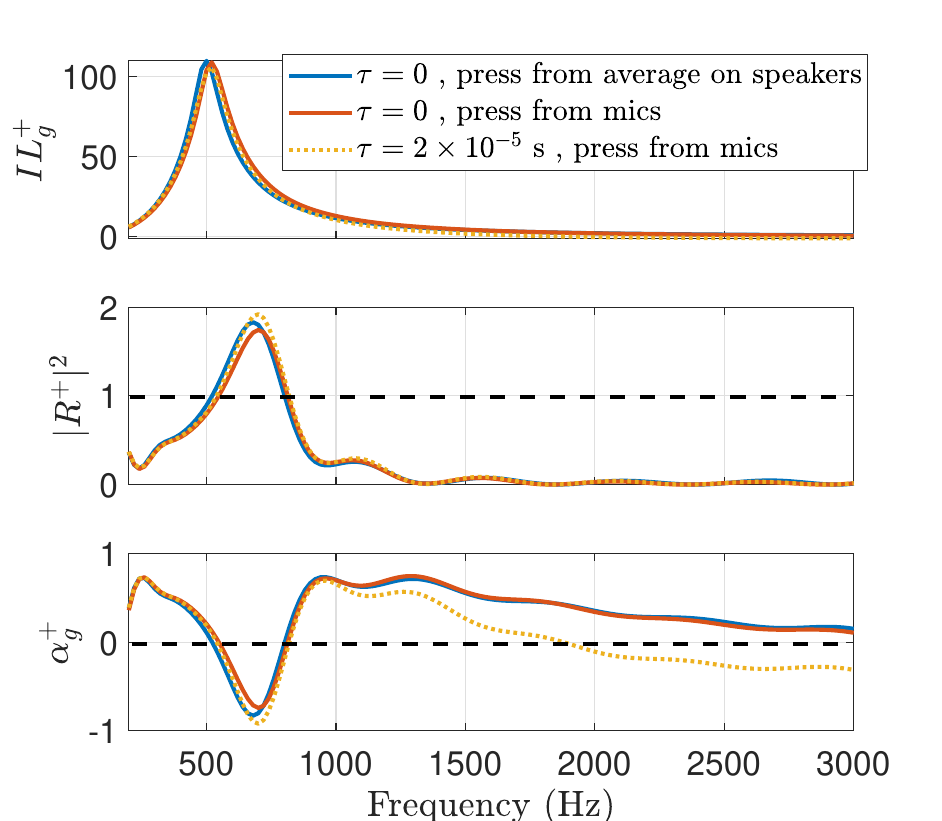}\\
	\caption{Scattering coefficients in a 3D waveguide of cross section width $h=0.05$ m lined with ABL ($\mu_M=\mu_K=0.5$, $R_d=\rho_0c_0$, $M_b=-1$), in case of pressure average evaluation on speakers and no delay (in blue), in case of pressure evaluated from microphones averaging (see Fig. \ref{fig:loudspeaker_with_NonLocalControl_scheme}) and no delay (in red), and in case of pressure evaluated from corner microphones and with time delay (in yellow).}
	\label{fig:SCATTplus_3D_0.05CellDim_6cellsPerLine_-1cac0_0.5muMmuK_1rho0c0Rat}
\end{figure}

In order to assess the effect of the pressure estimation from the 4 corner microphones on each ER, in case of ABL, in Fig. \ref{fig:SCATTplus_3D_0.05CellDim_6cellsPerLine_-1cac0_0.5muMmuK_1rho0c0Rat} we report the simulated scattering performances in a 3D waveguide, when $\hat{p}$ and $\hat{\partial}_x p$ are retrieved from the average values on each ER disk (solid blue), or when $\hat{p}$ and $\hat{\partial}_x p$ are obtained from the 4 corner microphones (solid red). In dotted yellow, we also report the simulations results when a time delay $\tau=2\times10^{-5}$ seconds is considered in the controller. Observe that, employing the corner microphones to estimate $\hat{p}$ and $\hat{\partial}_x p$ slightly affects the scattering performances around resonance and reduce the high-frequency passivity. The addition of a time delay in the control algorithm strongly affects the acoustical passivity at high frequencies, with $\alpha_g^+$ and $IL_g^+$ becoming negative, as expected \cite{de2022effect}.

\end{appendices}

\FloatBarrier

	\bibliographystyle{model1-num-names}

\end{document}